\documentclass[a4paper,twocolumn,10pt]{quantumarticle}
\pdfoutput=1
\usepackage[english]{babel}
\usepackage[T1]{fontenc}
\usepackage{hyperref}
\usepackage{comment}
\usepackage{bookmark}
\usepackage{xcolor}
\usepackage{braket}
\usepackage{tikz}
\usepackage{amsmath}
\usepackage{amssymb}
\usetikzlibrary{quantikz}
\usepackage{lipsum}
\usepackage{rotating}
\usepackage[export]{adjustbox}

\usepackage{booktabs}
\usepackage{subcaption}
\usepackage{graphicx}
\usepackage{pifont}
\newcommand{\cmark}{\ding{51}} 
\newcommand{\xmark}{\ding{55}} 

\begin{document}

\title{Resource Estimation via Efficient Compilation of Key Quantum Primitives
}
\author{Colin Campbell}
\email{colin.campbell@infleqtion.com}
\affiliation{Infleqtion, Chicago, IL, USA}
\author{Rich Rines}
\affiliation{Infleqtion, Chicago, IL, USA}
\author{Victory Omole}
\affiliation{Infleqtion, Chicago, IL, USA}
\author{Tina Oberoi}
\affiliation{University of Chicago, Chicago, IL, USA}
\author{Palash Goiporia}
\affiliation{Infleqtion, Chicago, IL, USA}
\author{Rayat Roy}
\affiliation{University of Chicago, Chicago, IL, USA}
\author{R. Peyton Cline}
\affiliation{Infleqtion, Louisville, CO, USA}
\author{Eric B. Jones}
\affiliation{Infleqtion, Louisville, CO, USA}
\author{Teague Tomesh}
\affiliation{Infleqtion, Chicago, IL, USA}
\email{teague.tomesh@infleqtion.com}

\maketitle

\begin{abstract}
    Resource estimation is a significant challenge in evaluating fault tolerant quantum computers. Existing approaches often rely on either fixed architectural assumptions or coarse analytical models that fail to capture the interaction between hardware constraints and circuit compilation.
    This challenge is particularly acute for neutral atom quantum computers, where architectural features such as atom movement, measurement zones, and multi-species arrays introduce a broad design space for implementing fault tolerant computation.
    Addressing the need for a tighter feedback loop between hardware design and practical application development,
    we present a compilation-driven framework for quantum resource estimation that translates arbitrary quantum circuits into logical primitive operations with known physical resource costs. 
    This framework allows for easily configurable hardware assumptions that enable rapid comparison of different architectural design choices.    
    We apply our approach to two early fault tolerant quantum simulation and optimization workloads, assuming the use of the surface code, revealing several architectural trends.
    While the production of magic states continues to be the dominant source of overhead for these benchmarks, access to movement can save time on cultivation and important transversal gates.
    As problem size grows, routing and qubit movement become dominant bottlenecks, highlighting the need for movement-aware compiler optimizations and frugal routing strategies. Finally, our results suggest that neutral atom architectures combining dual-species arrays with controlled qubit movement offer a promising path toward near-term advantage on fault tolerant devices.
\end{abstract}
\begin{figure*}[th]
    \centering
    \includegraphics[width=\textwidth]{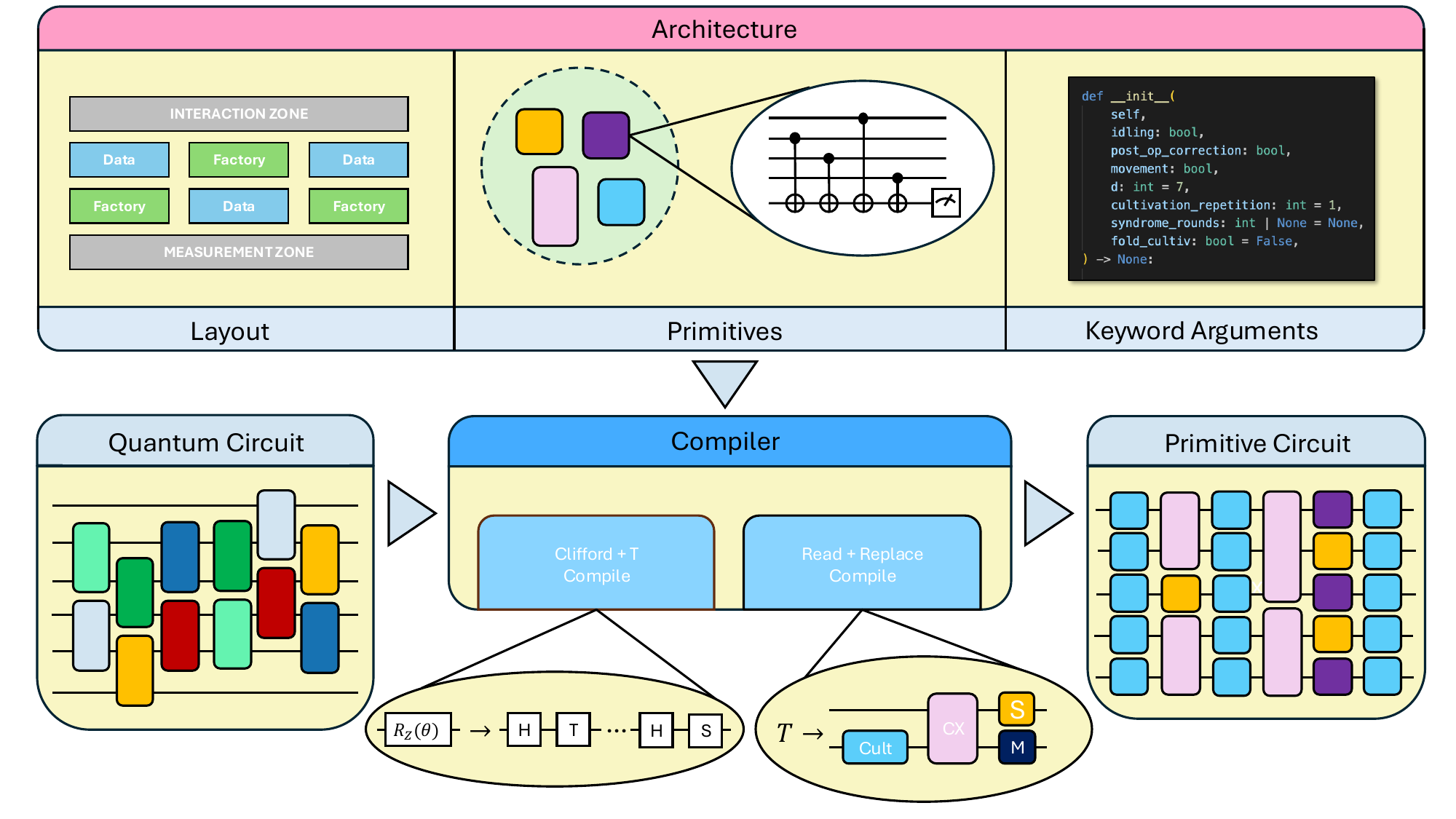}
    \caption{Overview of the resource estimation pipeline. The compiler takes as input an arbitrary quantum circuit and an Architecture class. The Architecture stores information including a universal fault tolerant set of logical operations called primitives, a logical layout of qubits, and a set of keyword arguments to further specify assumptions. The Compiler uses a two-stage approach to produce a circuit composed of primitives amenable to resource estimation. The first stage compiles to Clifford gates and single qubit rotations approximated by discrete gates. The second stage iteratively applies replacement rules until only primitives remain. Since the Architecture also contains pre-compiled physical decompositions of primitives, the resources can be estimated according to the sum of primitives along the output circuit's critical path.}
    \label{fig:overview}
\end{figure*}

\begin{figure*}[th]
    \centering
    \includegraphics[width=.8\linewidth]{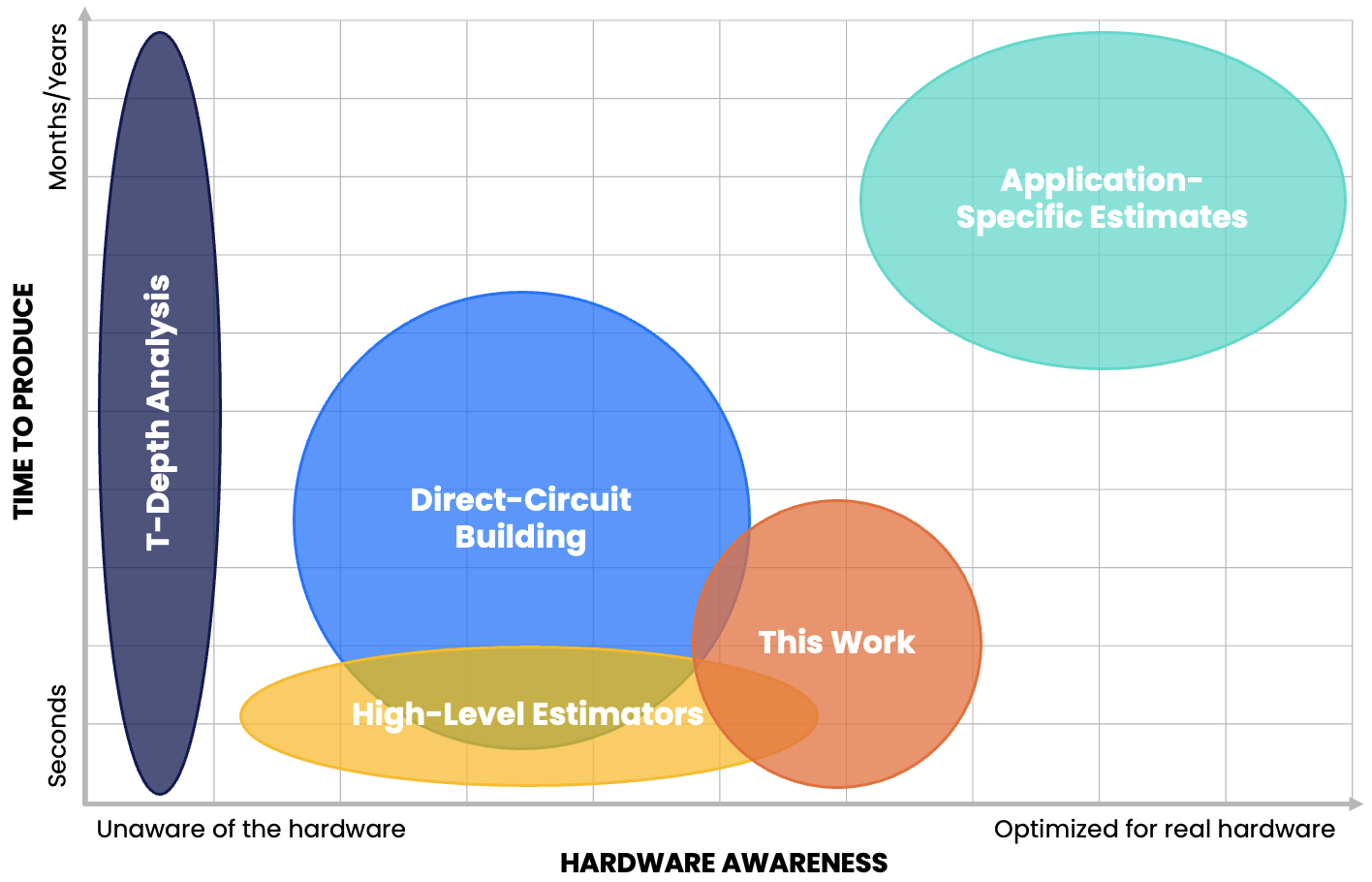}
    \caption{Typical space of tradeoffs between formula-based hardware-agnostic resource estimates and application specific estimates optimized for real hardware. Thorough analyses dedicated to a single application or circuit, taking months or years to produce, are not easily reconfigurable under new assumptions. On the other hand, writing down logical circuits or performing scaling studies without knowledge of the underlying hardware can be rapidly re-analyzed under different assumptions, but they often lack the necessary precision need to understand the challenges of running on real hardware. Our work aims to bridge this gap in precision by making the assumptions both reasonable and reconfigurable and the inputs arbitrary quantum circuits, balancing the need for the highest quality estimates against the need for rapid answers facilitating architectural comparisons.}
    \label{fig:tradeoff}
\end{figure*}

\section{Introduction}
\label{sec:intro}
\quad Quantum resource estimation appears to be a deceptively simple task; nevertheless, it remains one of the central challenges in assessing the practical potential of quantum computation. 
Many quantum applications in optimization~\cite{shaydulin2021qaoakit}, chemistry~\cite{mcclean2020openfermion}, and machine learning~\cite{bergholm2022pennylaneautomaticdifferentiationhybrid} already possess out-of-the box circuit representations, and a growing ecosystem of additional software tools has made experimenting with these quantum programs much easier~\cite{qiskit2024, Cirq_Developers_2025}.
However, the rapidly expanding capabilities of software tools and hardware platforms raise several natural questions. How much time will it take to run a useful application on a real quantum computer? How many qubits will the quantum computer need to run it? What gate fidelity must it have? 
In practice, answering these questions remains difficult because the direct execution of useful quantum applications has so far been limited by both noise and system size.
While different claims of quantum advantage~\cite{doi:10.1126/science.ado6285} or quantum supremacy~\cite{arute2019quantum} have been reported, present-day systems have not yet enabled commercially relevant breakthroughs such as breaking the RSA cryptosystem~\cite{Rivest1978-zo}, discovering room temperature superconductors, or replacing classical approaches for solving hard optimization problems.
Considerable progress has been made in compiling and optimizing circuits for near-term quantum hardware~\cite{campbell2023superstaq,Carvalho_2021}, and impressive demonstrations of steady hardware improvements provide evidence that quantum computers are on the path towards practical relevance~\cite{mayo2026benchmarkingquantumcomputersprotocols, bluvstein2025logical, rines2025demonstration, google2025quantum, dasu2026computing}. Nevertheless, realizing large, advantage-scale quantum applications will likely require fault tolerance through the use of quantum error correction (QEC)~\cite{Preskill1998-rk, Terhal2015-xx, Campbell2017-xi}.

Recognizing the importance of this additional layer of complexity has set off a gold rush of theoretical work across a wild west of QEC schemes. While some approaches have matured over decades~\cite{Kitaev_2003, bravyi1998quantumcodeslatticeboundary, Horsman_2012, Fowler_2012, litinski2019gameofsurfacecodes, gidney2024magicstatecultivationgrowing}, some are brand new as of the writing of this paper~\cite{koh2026entanglinglogicalqubitsphysical}.  
The added complexity can be overwhelming when trying to select the best combination of a specific quantum application, a realistic hardware modality, and a well-defined QEC scheme. For some modalities the challenge of solving this problem is eased by having fixed constraints on their gates and connectivity. Superconducting architectures, for example, boast fast gate speeds and high gate fidelities~\cite{mayo2026benchmarkingquantumcomputersprotocols}, but are typically limited in connectivity between physical qubits and pay a high price for idling due to their short coherence times. Other modalities, like neutral atoms, have many proposed implementations~\cite{bluvstein2022, Li_2025, radnaev2024universal, Anand2024}. The use of atom motion to facilitate higher physical connectivity~\cite{rines2025demonstration, bluvstein2024logical}, the potential for programmable multi-qubit gates~\cite{PhysRevApplied.23.054074}, and the utilization of inter-species quantum gates~\cite{Anand2024} 
are all examples of design decisions that can define trajectories for the future of neutral atom quantum computing. However, implementation and evaluation of these and other design decisions requires substantial effort in theoretical modeling and in device engineering. Rapidly estimating the impact of these design decisions on total quantum resources for applications of interest is therefore critical for building the first generation of neutral atom quantum computers that can achieve quantum advantage in the near term.

To this end, we present an open source package that estimates the quantum resources for circuits under a reconfigurable set of assumptions with a particular focus on neutral atom modalities. Our work also functions as a synthesis of many techniques in the neutral atom and QEC literature. This synthesis is a necessary component of describing the process of fault tolerant resource estimation. Our primary goal is to quantify key metrics like circuit time and physical qubits based on reasonable and malleable sets of assumptions, which will allow us to evaluate rapidly which decisions have the highest impact on the key metrics. We provide a small set of operations called \textit{primitives} to build lower level circuit representations that clearly show the impact of different design decisions and determine their costs as a function of current and future gate speeds. These building blocks utilize the benefits of the heavy circuit compilation strategies developed to run Noisy Intermediate Scale Quantum (NISQ) circuits on early quantum computers. As another contribution, we assess the viability of magic state cultivation as a method to bridge the gap between small noisy systems and large error corrected ones.

The remainder of the paper is laid out as follows. In Section~\ref{sec:background} we provide a background on resource estimation for fault tolerant quantum computation, list alternative tools, and discuss the strengths and weaknesses of the prior art. Section~\ref{sec:core-components} delineates the core components of our model, describes its configurable parameters, and provides justification for the choices. Section~\ref{subsec:ft-compilation} details the full pipeline taking an abstract quantum circuit all the way to the most important quantum resources. Section~\ref{sec:experiments} demonstrates the utility of the tool through several example architectural comparisons relevant to neutral atoms. Finally, Section~\ref{sec:conclusion} summarizes key insights and conclusions, highlighting the importance of resource estimation tools as part of the larger ecosystem of quantum resource estimation software and the importance of cultivating this ecosystem to usher in the era of fault tolerant quantum computing.

\section{Background on Quantum Resource Estimation}
\label{sec:background}
\quad Throughout the era of NISQ computing~\cite{preskill2018quantum}, executing hardware experiments has been a balancing act between running circuits deep enough to display quantum properties and reducing the circuit complexity enough to extract signal from a noisy device. Understanding this delicate balance has led to a number of valuable innovations in circuit compilation~\cite{campbell2023superstaq, murali2019noise, wang2022quantumnas}, random circuit sampling~\cite{arute2019quantum, zhong2020quantum, madsen2022quantum}, and more efficient resource management~\cite{ding2020square, sahay2023high}.

\subsection{Resource Estimation with Formulas}
\label{subsec:formulas}
\quad Transitioning from physical qubits to logical qubits adds a considerable amount of complexity to the task of resource estimation. QEC codes are typically characterized by the notation $\left[\left[n, k, d\right]\right]$, where $n$ is the number of physical qubits, $k$ is the number of logical qubits, and $d$ is the code distance. The code distance $d$ determines the code's resilience against errors, and a higher code distance exponentially suppresses the set of errors that can enter the computation undetected. In practice, the distance provides a fidelity for logical operations and can be thought of as boosting the physical fidelities to a much higher level at the cost of greater overhead. The literature often uses threshold plots or formulas to illustrate the relationship between physical noise, code distance, and logical error rate. For codes where the threshold is known and given a basic physical error rate, one may derive the required code distance to perform enough circuit operations to remain below a desired noise threshold. Furthermore, with $d$ derived from the number of circuit operations, and $k$ loosely given by the number of logical qubits in the circuit one can also derive the number of physical qubits needed.

This approach to resource estimation is the most straightforward and is often the basic approach taken in work interested in asymptotics. Here we work through an oversimplified, but illuminating, toy example of resource estimation for Shor's Algorithm~\cite{shor1994algorithms} using the rotated surface code. The rotated surface code is characterized by $\left[\left[d^2,1,d\right]\right]$
, and has a characteristic fidelity of 
\begin{equation}
    f=1-0.03\left(\frac{p}{p_{th}}\right)^{\frac{d+1}{2}} \label{eq1},
\end{equation}
with a threshold $p_{th}=0.0057$~\cite{Fowler_2012}. Resource estimates typically assume the error rate $p=0.001$, which leads to a logical error rate (LER) of
\begin{equation}
    0.03\left(0.17\right)^{\left(\frac{d+1}{2}\right)}. \label{eq2}
\end{equation} 
Note that as $d$ increases, the LER decreases exponentially, while the number of physical qubits $n$ increases quadratically.

Table I of Ref.~\cite{Fowler_2012} provides a basis for the logical qubit count, Toffoli count, and circuit depth for Shor's Algorithm. Using the formulas therein, we can make a simplified resource estimate to factor RSA integers. To factor an $N$-bit number, the table suggests $2N$ qubits and $4N^3$ gates. Plugging in $N=2048$, we get $4096$ logical qubits and $3.4\times 10^{11}$ operations. To include potential other gates and get a round number, we can set a target LER of $10^{-12}$. The smallest odd $d$ for which Eq. \eqref{eq2} is below the LER is $27$. Therefore we need $729$ physical qubits for each of the $4096$ logical qubits, resulting in a total of $3.0\times 10^6$ physical qubits to run Shor's Algorithm. As we will show, the simplicity of this approach hides a significant degree of complexity that underpins the study of QEC as a subject. Nevertheless, the key takeaway is that by sacrificing precision, we can estimate the required resources much faster than going into the details of performing a fully fault tolerant compilation.

While the simplified estimate is fast and straightforward, it neglects one of the most important components of a good estimate--the circuit runtime. For the rotated surface code a simplifying assumption is to use the notion of ``cycles''. The term ``cycles'' refers to the time required to probe the quantum system to determine if and where an error has occurred in the computation, meaning that one moment of quantum logic corresponds to $d$ cycles. To simplify further, the corresponding cycle time is assumed to be one $1\mu s$ for superconductors~\cite{litinski2019gameofsurfacecodes} and $1ms$ for atomic qubits~\cite{sunami2025}. We return to the simplified example, which has a circuit depth of $10^{11}$. We also assume that $\sim10$ trials required to find a valid factor~\cite{erka-period-finding}. Therefore, we can estimate the total time as
\begin{align*}
    \text{Total Time} &= 10^{11} \frac{\text{ Moments}}{\text{Trial}} \times 27 \frac{\text{Cycles}}{\text{Moment}} \\ & \times 10^{-6}\frac{\text{Seconds}}{\text{Cycles}}
    \times 10\text{ Trials} \\ & \approx 10\text{ Months}.
\end{align*}
This estimate lacks precision because it does not consider the details of implementation and just relies on simple formulas. More modern predictions of the quantum requirements to run Shor's Algorithm estimate far fewer resources because they consider more implementation details.

\subsection{Incorporating Fault Tolerant Logic}
\label{subsec:FT-logic}
\quad The first step toward a more precise resource estimation is understanding how the logical gates are implemented. The $\left[\left[n, k, d\right]\right]$ description of a code does not necessarily provide a prescription for implementing logical gates. At the time of writing, Ref.~\cite{kasai2026breakingorthogonalitybarrierquantum} is an example of a promising QEC code that does not yet have a practical gate implementation. Code discovery is half the battle of developing a full QEC framework because protecting information is often a separate task from performing computation on the information. Complicating matters further, the Eastin-Knill theorem~\cite{eastin2009restrictions} forbids the existence of any QEC code admitting a universal transversal gateset. As a consequence, fault tolerant quantum computation must rely on additional non-transversal resources outside the intrinsic parameters of the code, most commonly in the form of distilled ``magic'' states~\cite{howard2014contextuality, haug2023scalable}. 

There are thus two important components that must be understood in addition to a code's $\left[\left[n,k,d\right]\right]$ representation: the implementation of logical gates and the production and distribution of magic. The tension between the many possible ways to physically implement logical gates in hardware and the many possible gatesets for QEC codes result in resource estimation approaches that fall into one of two camps. Figure~\ref{fig:tradeoff} illustrates the space of tradeoffs for different approaches to resource estimation. Many approaches tend to be either quick and hardware agnostic or lengthy research questions for specific applications.

On the one hand, applications and algorithms developers may define a quantum circuit in terms of one of the popular universal gatesets, such as Clifford + T, Clifford + Toffoli, or CNOT + 1Q gates. With such a gateset in hand, the circuit may be compiled carefully to the physical gateset. The resources can then be estimated according to formulas derived from the QEC code like Equation \eqref{eq1}, with more nuance included. In a similar way that CNOT count or CNOT depth is the metric of interest when considering NISQ circuits, fault tolerant resource estimates typically assume T to be the most expensive gate and report its count or depth. While resource estimates of this character are useful for establishing asymptotic scaling or facilitating high level comparisons between similar circuits, they often neglect the physical implementation details that are necessary to get an accurate count.
For example, recent work studying quantum optimization on 8-SAT problems~\cite{omanakuttan2025threshold} yields compelling resource estimates based primarily on formulaic gate-count models, though further accuracy may be achieved by accounting for key architectural constraints such as qubit routing.

On the other hand, experts in specific modalities can produce high effort resource estimates that come as a result of specifying nearly all involved parameters for a given quantum circuit to produce a highly precise estimate, often with an accompanying circuit to run on the available hardware. This end of the spectrum has great value for hardware teams that have few options for changing the characteristic parameters of their modality, but these estimates can become obsolete quickly when new techniques arise. For example, in the past two years, magic state cultivation has become a popular method for acquiring magic, and there have been several proposals for implementing it with a special focus on the architectural properties of neutral atoms~\cite{chen2025efficient, claes2025cultivating, sahay2025foldtransversalsurfacecodecultivation}. For this reason, there is practical value in developing software to explore the design space between application, QEC code, and hardware modality.

\subsection{Prior Work}
\label{subsec:prior-work}

\quad The most straightforward way to address this challenge is by having a compiler that compiles fault tolerant quantum circuits to hardware, like Ref.~\cite{yoder2025tourgrossmodularquantum}. While closely aligned with the goals outlined here, the tool is limited in its scope and tailored to superconducting architectures. Zapata's BenchQ~\cite{benchq} suffers from the same combination of narrow scope and hardware focus. Similarly, Refs.~\cite{litinski2019gameofsurfacecodes} and \cite{sunami2025} offer useful frameworks for addressing the challenge of compilation, but lack a software implementation to develop upon and are thus limited to the conditions they assume at the time of publication. Additionally, all of these approaches lack the ability to change the more qualitative assumptions---such as access to movement, degree of such access, method of magic acquisition, physical layout, physical gate speeds, etc. that are of the most practical concern to the ones implementing the design decisions in hardware.

At the time of writing, arguably the approach that most closely addresses the need we identify, while also being open source and freely available, is the Microsoft Azure Resource Estimator~\cite{beverland2022assessingrequirementsscalepractical}.
But with all these strengths, it lacks the lower level tools needed to compare important architectural questions involving the impact of movement, dual species, physical gate times, and layout of resource factories. Additionally, it relies on a unique programming language, Q\#, limiting its compatibility with other software tools and yielding a higher barrier to entry for quantum computing researchers. Additionally, the Microsoft Azure Resource Estimator is formula-based resource estimator, and therefore does not produce a lower level representation of the circuit, which is necessary to run the circuit fault tolerantly on real hardware. In this work, we aim to provide the tools and software to build realistic estimates that move closer to the highly detailed estimates without sacrificing speed and flexibility.

\begin{figure}[t]
    \centering
    \includegraphics[width=.48\textwidth]{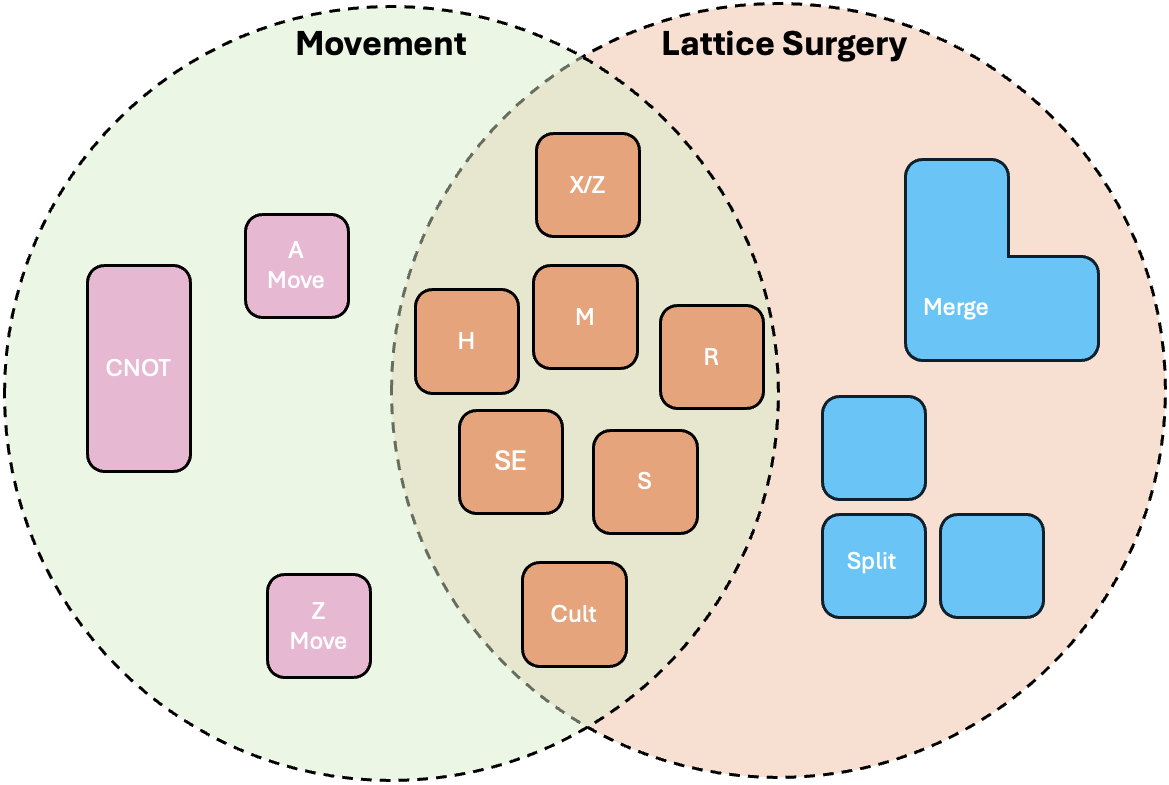}
    \caption{The relationship between sets of primitives for movement and lattice surgery. The resource costs will vary between Architectures and hardware implementations.}
    \label{fig:venn-diagram}
\end{figure}

\section{Core Components}
\label{sec:core-components}

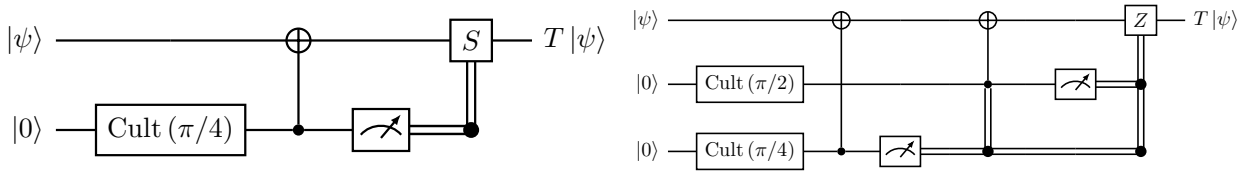
\begin{figure*}[th]
\centering
\begin{subfigure}[b]{.48\textwidth}
\centering
\resizebox{\linewidth}{!}{%
\begin{quantikz}
\lstick{$\ket{\psi}$} & \qw & \targ{} & \qw & \gate{S} & \qw \rstick{$T\ket{\psi}$} \\
\lstick{$\ket{0}$} & \gate{\text{Cult}\left(\pi/4\right)} & \ctrl{-1}  & \meter{} & \cw \cwbend{-1}
\end{quantikz}
}
\end{subfigure}
\begin{subfigure}[b]{.48\textwidth}
\centering
\resizebox{\linewidth}{!}{%
\begin{quantikz}
\lstick{$\ket{\psi}$} & \qw & \targ{} & \qw & \qw & \targ{} & \qw & \qw & \gate{Z} & \qw \rstick{$T\ket{\psi}$} \\
\lstick{$\ket{0}$} & \gate{\text{Cult}\left(\pi/2\right)} & \qw & \qw & \qw & \ctrl{-1} & \qw & \meter{} & \cwbend{-1} \\
\lstick{$\ket{0}$} & \gate{\text{Cult}\left(\pi/4\right)} & \ctrl{-2} & \meter{} & \cw & \cwbend{-1} & \cw & \cw & \cwbend{-1}
\end{quantikz}
}
\end{subfigure}
\caption{(a) T gate teleportation circuit with access to transversal S gate. (b) T gate teleportation circuit without access to a transversal S gate. The semi-classical CNOT represents performing a quantum CNOT if the classical wire is in the $|1\rangle$ state.}
\label{fig:teleportation-circuits}
\end{figure*}

\quad In this section we lay out many of the key components underpinning our resource estimation software, its fundamental assumptions, and the opportunities our framework provides.
The resource estimation pipeline begins with an input quantum circuit, typically written using \texttt{cirq} and accompanied by a top-level error tolerance for the whole circuit. For other quantum SDKs we refer users to open source conversion libraries like Ref.~\cite{campbell2023superstaq}. Starting with arbitrary quantum circuits allows our analysis to span the large space of quantum applications describable by the circuit model. While some modalities do process information in ways other than the circuit model~\cite{djidjev2018efficientcombinatorialoptimizationusing, Zhang2026-ih, Aghaee_Rad2025-wh}, QEC theory has largely developed within the assumption of a quantum circuit model.

Having a larger space of input quantum circuits also allows our work to go beyond the traditional approach to resource estimation. Many estimates begin with a Clifford + T circuit, which neglects the effects of approximation error for continuous rotations. We explicitly include circuits that can vary significantly between problem instances and contain many continuous angle rotations that must be compiled to Clifford + T first.

\subsection{QEC Assumptions}
\label{subsec:baseline-assumptions}
\quad Our model uses the rotated surface code as the default QEC scheme. While each logical qubit has a steep cost of a quadratic number of physical qubits, the rotated surface code boasts one of the highest thresholds, making it well-suited for early fault tolerant analysis. The rotated surface code also has one of the most extensive bodies of literature surrounding it of any of today's QEC schemes, including a complete set of gate implementation details, which are critical to obtaining high quality resource estimates.
In contrast, many exotic Quantum Low Density Parity Check (QLDPC) schemes exist with potentially better QEC properties, but none has been subject to the level of analysis and scrutiny of the rotated surface code.

While our approach focuses closely on this QEC scheme, it is not inherently limited to one code. The purpose of the framework is to show how having well-defined implementations opens the door to architectural comparisons and optimization opportunities.  Our approach uses those well-defined, physical-level circuit implementations to generate more accurate resource estimates, and the abstraction they enable allows for easy reconfigurability to study other quantum architectures and QEC codes. As more codes with more gatesets become available for neutral atom architectures, our approach to resource estimation can naturally fold them into its framework. 

We decide to focus on magic state cultivation~\cite{gidney2024magicstatecultivationgrowing} as our source of ``magic''~\cite{Oliviero_2022} in this work 
. The more traditional approach to preparing resource states is the use of magic state distillation~\cite{bravyi-distillation}, but we choose cultivation as our primary source of magic for several reasons.

First, cultivation offers low logical error rates relatively cheaply. Although the cost of distillation has decreased significantly over time~\cite{Litinski:2019uvg}, it is still a costly way to generate magic in fault tolerant quantum circuits. In contrast, cultivation's logical error rates between $10^{-6}$ and $10^{-9}$ are much lower than the physical noise assumed in the system and can potentially run applications in the advantage regime~\cite{jamet2025anderson, jones2025dynamic}.

Magic state cultivation has also benefited from tremendous work by the neutral atom community. Through the use of atom rearrangement, that enable physical gates beyond nearest neighbors, neutral atom platforms can implement higher fidelity cultivation at a fraction of the cost~\cite{sahay2025foldtransversalsurfacecodecultivation}. Appendix~\ref{app:cultivation} discusses three cultivation strategies in more detail, but the majority of this work assumes one of two approaches (Table~\ref{tab:arch-assumptions}).

Lastly, magic state cultivation's footprint of a single logical qubit makes it particularly appealing for routing and scheduling resource distribution in compiled fault tolerant circuits.

Notably, deciding to focus on cultivation does not restrict this approach to the rotated surface code in the long term because no QEC scheme can produce a universal gateset on its own. All methods will require generation and allocation of similar resources as a key subroutine. Our flexible framework for completing this task has direct application to any QEC scheme being used to fault tolerantly compile an input quantum circuit.

\subsection{The Architecture Class}
\label{subsec:architecture}
\quad The \textit{Architecture} class is the key component of our approach to fault tolerant resource estimation (see Figure~\ref{fig:overview}). It is composed of three primary components: a set of primitives, a logical qubit layout, and a set of keyword arguments that are used to augment the implementation details of the set of primitives.

\subsubsection{Primitives}
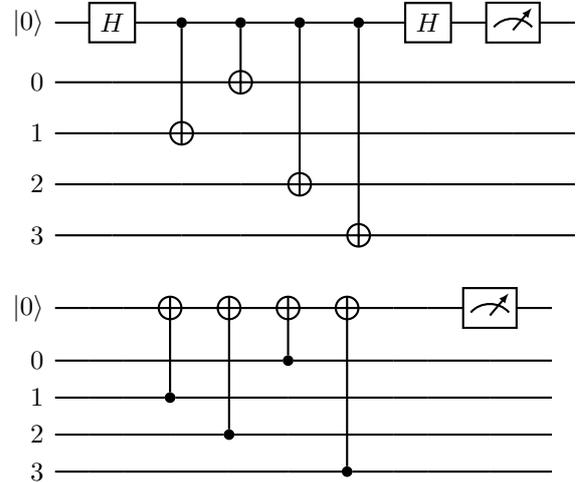
\begin{figure}[t]
\begin{quantikz}[row sep=0.35cm, column sep=0.45cm]
\lstick{$\ket{0}$} & \gate{H}
  & \ctrl{2} & \ctrl{1} & \ctrl{3} & \ctrl{4}
  & \gate{H} & \meter{} & \qw \\
\lstick{$0$} & \qw
  & \qw      & \targ{}      & \qw  & \qw
  & \qw      & \qw      & \qw \\
\lstick{$1$} & \qw
  & \targ{}      & \qw      & \qw      & \qw
  & \qw      & \qw      & \qw \\
\lstick{$2$} & \qw
  & \qw  & \qw      & \targ{}      & \qw
  & \qw      & \qw      & \qw \\
\lstick{$3$} & \qw
  & \qw      & \qw  & \qw      & \targ{}
  & \qw      & \qw      & \qw
\end{quantikz}

\vspace{0.8em}

\begin{quantikz}[row sep=0.35cm, column sep=0.45cm]
\lstick{$\ket{0}$} & \qw & \qw
  & \targ{}  & \targ{}  & \targ{}  & \targ{} & \qw & \qw
  & \meter{} & \qw \\
\lstick{$0$} & \qw & \qw
  & \qw      & \qw      & \ctrl{-1}      & \qw
  & \qw      & \qw & \qw & \qw\\
\lstick{$1$} & \qw & \qw
  & \ctrl{-2}      & \qw & \qw      & \qw
  & \qw      & \qw & \qw & \qw\\
\lstick{$2$} & \qw & \qw
  & \qw & \ctrl{-3}      & \qw      & \qw
  & \qw      & \qw & \qw & \qw \\
\lstick{$3$} & \qw & \qw
  & \qw      & \qw      & \qw & \ctrl{-4}
  & \qw      & \qw & \qw & \qw
\end{quantikz}
\caption{Example stabilizer measurement circuits for $X$ (Top) and $Z$ (Bottom) in the rotated surface code. The Syndrome Extraction primitive is the measurement of all stabilizers across a full code patch.}
\label{fig:se-circuits}
\end{figure}

\quad The term \textit{primitive} describes an element of a fault tolerant instruction set.
These instructions can include transversal logical operations, methods of generating magic, operations for facilitating entanglement, necessary overhead to keep a computation fault tolerant, and any other core operation needed for a specific QEC scheme. The operations should be the size of NISQ circuits and be relatively hardware agnostic. For example, consider a stabilizer measurement circuit for the surface code in Figure~\ref{fig:se-circuits}.

A full round of Syndrome Extraction applies these short circuits across the whole qubit patch. These circuits are shallow, hardware agnostic, and critical to maintaining fault tolerance, so they are a natural choice for a primitive.

We consider two sets of primitives in this work--one based on transversal operations enabled by movement and one based on lattice surgery. As shown in Figure~\ref{fig:venn-diagram}, they have a high degree of overlap, so we begin by describing the elements they have in common.

First, as mentioned above, we include Syndrome Extraction as a core component of both sets. We also include the Pauli gates as fundamental components for quantum operations; however, for the purpose of resource estimation, their cost can typically be ignored because they can be implemented virtually. We give both sets access to the Hadamard (H) gate because although its physical implementation is significantly different between architectures, both implementations look similar to performing a known intermediate-scale circuit. Specifically, we use Ref.~\cite{Geh_r_2024} for our lattice surgery implementation of the H gate. For movement architectures we use local single qubit gates, followed by a Movement operation to represent the qubit re-permutation returning the stabilizers to their correct configuration. We include logical Measurement as a primitive because it is critical to teleporting magic from outside the code, and it has the potential to unlock spacetime tradeoffs~\cite{hall2026quantumdepthcompressionlocal} that may be crucial to neutral atoms. We model the operation as a physical measurement across all physical qubits in the code patch. Similarly, we include a Reset primitive as a patch-wide physical reset operation. The final primitive that is common between all architectures in this paper is Cultivation (``Cult-T'' or ``Cult-S'' depending on the state). This operation generates resource states like $\ket{T}$ and $\ket{S}$ that can be teleported into the main circuit. They are typically also the most time consuming operations (See Table~\ref{tab:primitive-times}).

The main driver of the lattice surgery primitives is the expectation that many hardware implementations like superconducting will be limited to a nearest-neighbor connectivity that makes fault tolerant logic more challenging. To address this challenge, we include Merge and Split operations in the set of lattice surgery primitives. For our ``Superconducting Sea of Qubits'' model, these operations use logical patches of ancilla to distribute entanglement between logical computational qubits. A description of the placement of these qubits and their assigned roles can be found in the next section. 

Architectures with access to movement can partially overcome the limitations of lattice surgery. Though it needs to be balanced against potential re-cooling overhead, physical motion of atoms allows access to several transversal operations. Ref.~\cite{chen2024transversallogicalcliffordgates} gives a prescription for logical S gates requiring intra-patch moves. Physical H on a code patch yields a logical H if a qubit reordering is available~\cite{Geh_r_2024}. The CNOT gate can also be implemented transversally by applying entangling gates to matching data qubits in pairs of code patches~\cite{PRXQuantum.6.020326}. Thus, we add each of these operations to the movement set of primitives. Since we are interested in assessing the cost of movement and comparing it between architectures, we elect not to have one ``Movement'' primitive. Instead, we consider logical movement of two types: one for zone moves (Z-Move) like those in Ref.~\cite{harvard-arch} and one for moves through alleyways in the qubit array (A-Move) like those in Ref.~\cite{rines2025demonstration}. Typical times for each of these moves within our model can be found in Table~\ref{tab:primitive-times}, and Appendix~\ref{app:gate-speeds} gives more detail on the motivations for choices.

Within the Architecture, the set of primitives serves three purposes. First, it provides a clear instruction set architecture that defines how the quantum computer modeled under this framework will perform computation fault tolerantly. The quantum computer can be thought of as completing many of these smaller, simpler sub-circuits in sequence (with corrections and feed-forward logic) to achieve the larger computation. While the number may be great, the principle of using these operations serves as the basis for the fault tolerant compiler framework in Section~\ref{subsec:ft-compilation}. Second, having these small circuits simplifies the task of resource estimation significantly. The cost of a primitive can be calculated once and then be reused every time it is seen in a fault tolerant circuit. Third, isolating these operations, while keeping open the option to define new sets, allows for more efficient fault tolerant circuits through the careful optimization of a few key operations for the hardware instead of a complex complete fault tolerant circuit.

\subsubsection{Layout}

\quad The \textit{Layout} class is an extension of the connectivity map typically used to describe which entangling gates can be performed between physical qubits in NISQ machines. To expand the notion to logical architectures, we include a level of qubit assignment. One key component is specifying the number of factories, which allows us to experiment with spacetime tradeoffs in addition to qubit placement. We also expect to reuse many mapping and routing techniques that NISQ has utilized to improve circuit performance in the intermediate regime. Layouts take two main parameters: the number of T factories and, for lattice surgery Layouts, the number of S factories. These parameters in conjunction with the Layout generation strategy allow us to fully specify the positions and assignments of logical qubits.

For movement Architectures, logical qubit patches are assigned as either logical qubits or factory qubits. The qubits of each type all fill a logical array where the Manhattan distance mimics the physical distance needed to travel to move qubits to one another or to a zone. This layout method matches closely with Ref.~\cite{sunami2025}, but by using cultivation instead of distillation we have much more configurability in the layout of the logical array. While this model is useful, it is not always desirable for neutral atom architectures in practice. Ref.~\cite{harvard-arch} does not need a strict localization of qubits during storage because all two-qubit gates happen in the entangling zone. We offer more detail and alternative constructions in Appendix~\ref{app:layout-generation}.

\subsubsection{Extra Information}
\quad In our model we include several additional keyword arguments to complete the instantiation of an Architecture. The main parameter to specify is the code distance $d$, which will determine the qubit footprint, per logical patch and the depth of some of the primitives. The following section provides more details on choosing the code distance. Additionally, we include parameters for idling and Syndrome Extraction frequency, which are relevant to compiler passes in Section~\ref{subsec:ft-compilation}. They allow for experimentation of costs with more or less Syndrome Extraction on idling and active qubits. To explore the impact of correlated decoding~\cite{Zhou2025-ju, caincorrelated} we include a parameter to control the number of rounds of extraction within a single Syndrome Extraction primitive. Lastly, we include a parameter for turning on folded cultivation, which toggles between Refs.~\cite{gidney2024magicstatecultivationgrowing} and \cite{sahay2025foldtransversalsurfacecodecultivation} styles of cultivation. These parameters are usually set to their defaults for each Architecture (see Table~\ref{tab:arch-assumptions}), but having the ability to change them is useful for having a flexible approach to resource estimation.

\section{Resource Estimation Pipeline}
\label{subsec:ft-compilation}

\quad We now describe the pipeline for fault tolerant compilation, as shown in Figure~\ref{fig:overview}. As input the pipeline takes a quantum circuit and a desired circuit fidelity. The output is a lower level representation of the circuit in terms of primitives and the quantum resources. Our approach is a simple, step-by-step process. We also include an error sensitivity analysis to show potential benefits from doing some of the error budget management outside of the simple pipeline.
\subsection{Clifford + T Compilation}
\quad To begin, we assume a quantum circuit $C_0$ and requested maximum circuit error as the starting point of the pipeline. We allow $C_0$ to be any circuit expressed by a sequence of unitary gates (and possibly Measurement/Reset).

Clifford + T is a universal gateset that can approximate any quantum circuit to arbitrary precision. However, choosing that precision will impact the quantum resources required. Before making any approximations, we first convert $C_0$ into a gateset much closer to the discrete one. In this work, we use Clifford + $R_Z$, where $R_Z$ gates can take any continuous rotation angle $\theta$. We call the resulting circuit $C_1$. To perform this step, we implement several simple compiler passes to (a) break multi-qubit gates into single- and two- qubit gates, (b) decompose two-qubit gates into CNOT and single-qubit gates via the KAK decomposition~\cite{tucci2005introductioncartanskakdecomposition}, (c) decompose single qubit gates via Euler angles, and (d) merge/eject unnecessary gates.

To get from the Clifford + $R_Z$ circuit $C_1$ to the desired Clifford + T circuit $C_2$, we note that the Solovay-Kitaev Theorem~\cite{dawson2005solovaykitaevalgorithm} guarantees that we can approximate any continuous rotation operation efficiently. In particular, it is possible to do so with a sequence of $\mathcal{O}(\log\epsilon_{R_Z})$ H S, and T gates~\cite{ross2016optimalancillafreecliffordtapproximation}. Since we may have many $R_Z$ gates in an input circuit, we define $\epsilon_{R_Z}$ to be the error per $R_Z$ gate with the understanding that the approximation rule can be applied each time. We discuss methods for choosing $\epsilon_{R_Z}$ in the next section.

This two-part compilation approach is the first step in the pipeline, making it one of the most consequential in terms of downstream resource estimates. We use \texttt{cirq}~\cite{Cirq_Developers_2025} for the first stage and an open source python implementation called \texttt{pygridsynth}~\cite{pygridsynth} for the second. In the future, a dedicated Clifford + T compiler would improve the precision of our resource estimates by implementing optimizations that are not otherwise unavailable when the two stages are left separate. For example, combining single qubit gates into a single unitary approximated by one string of gates would significantly more efficient than doing two rotation syntheses for different Euler angles. One advantage we do see in this approach is the ability to separate Clifford and non-Clifford segments of the circuit early on in the process. This information can be important for hardware practitioners interested in the ratio between these gate types and can be useful in setting expectations for potential space-time tradeoffs. Finding high-quality logical circuits for specific gatesets is still an active field and one of the main facilitators of NISQ experiments.

\begin{figure}
    \centering
    \includegraphics[width=1\linewidth]{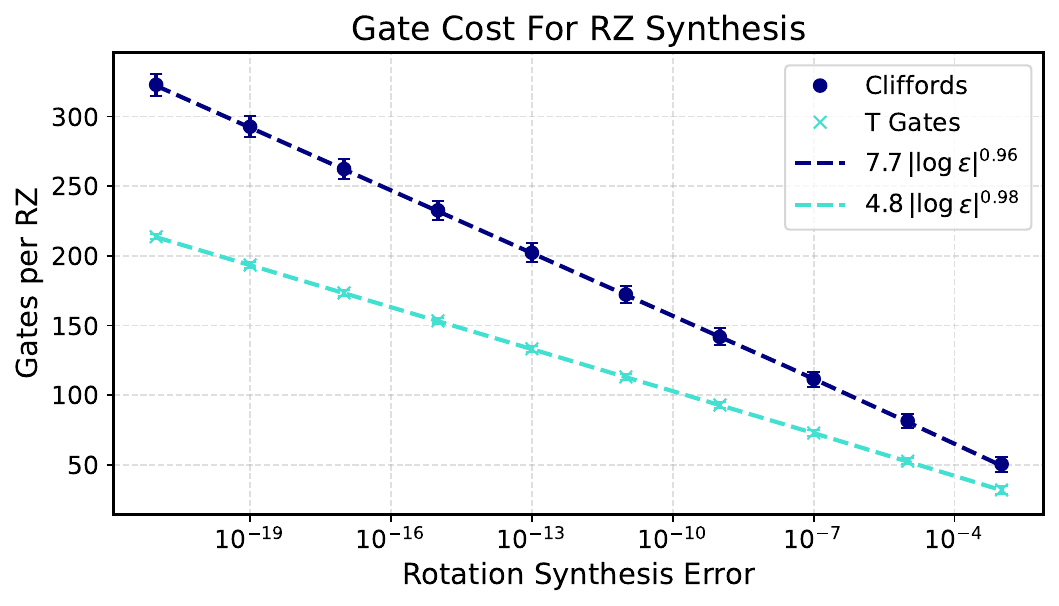}
    \caption{Gates used to approximate arbitrary $R_Z$ rotations via Ref \cite{ross2016optimalancillafreecliffordtapproximation}, broken down by Clifford and non-Clifford. Fit parameters can be used in for approximately optimizing overhead from synthesizing rotations against overhead from error correction.}
    \label{fig:synth_plot}
\end{figure}
\begin{figure*}[ht]
    \centering
    \begin{tabular}{@{}c|c@{}}
    \includegraphics[width=.48\linewidth]{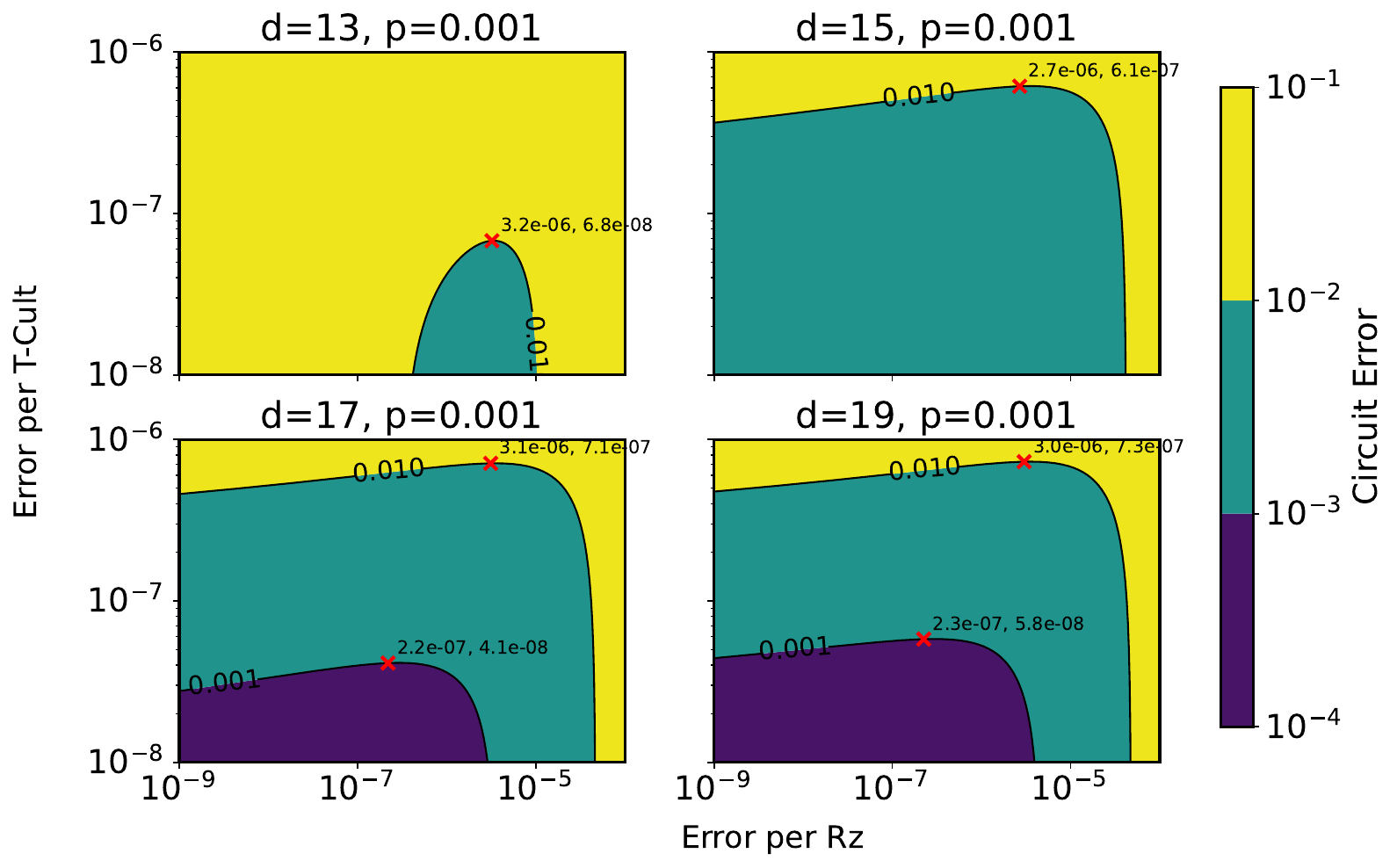} & 
    \includegraphics[width=.48\linewidth]{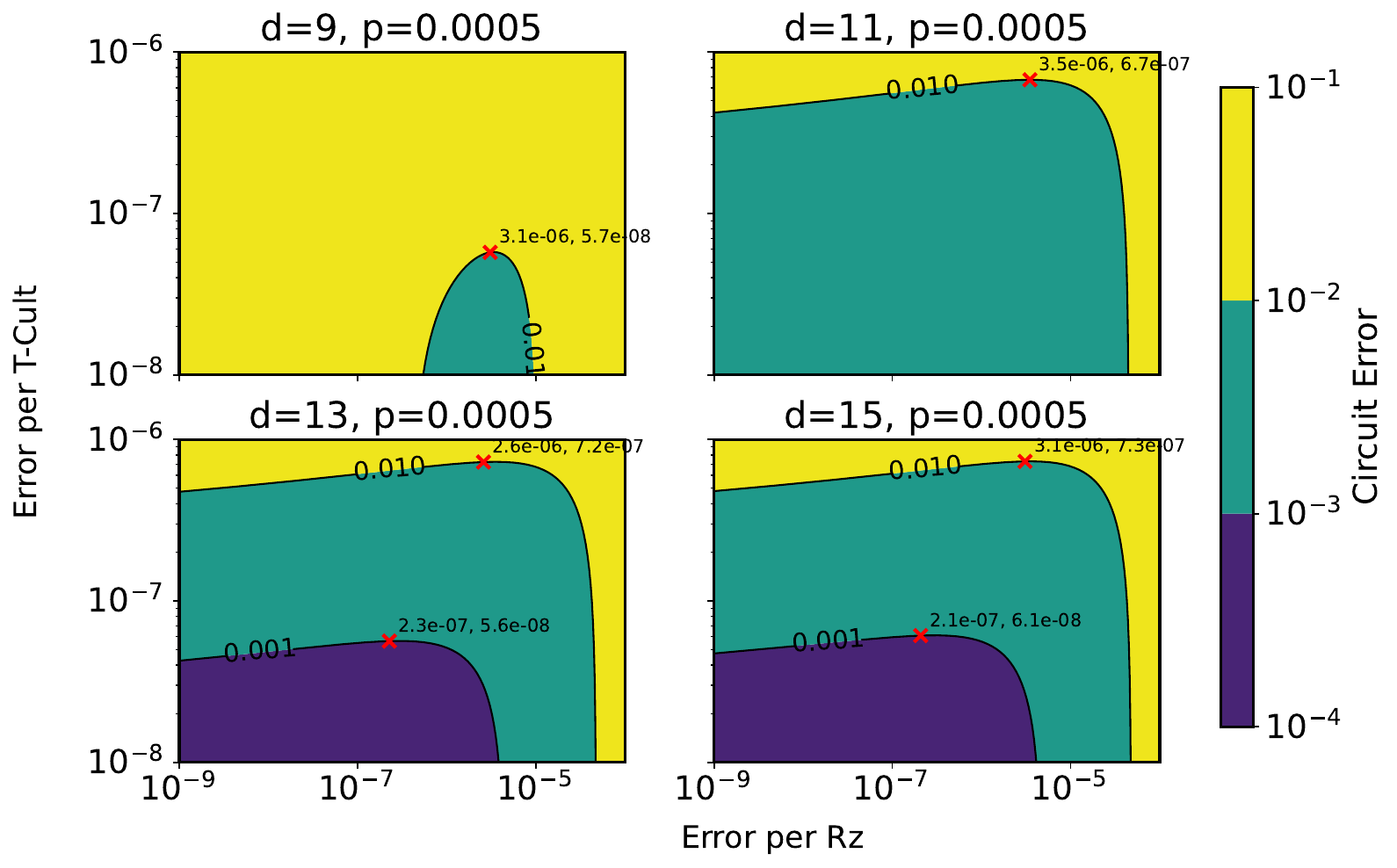} \\ \hline
    \includegraphics[width=.48\linewidth]{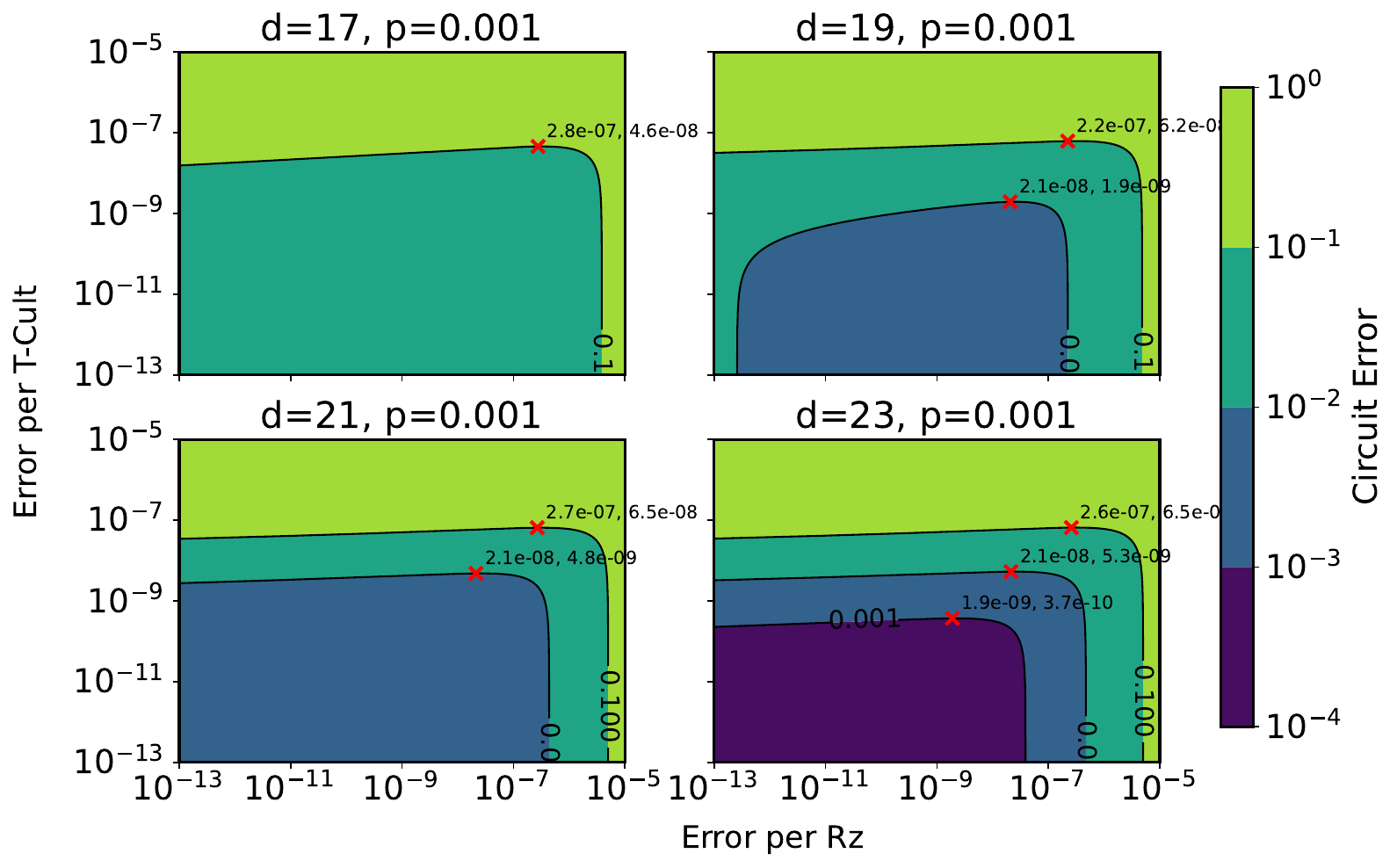} &
    \includegraphics[width=.48\linewidth]{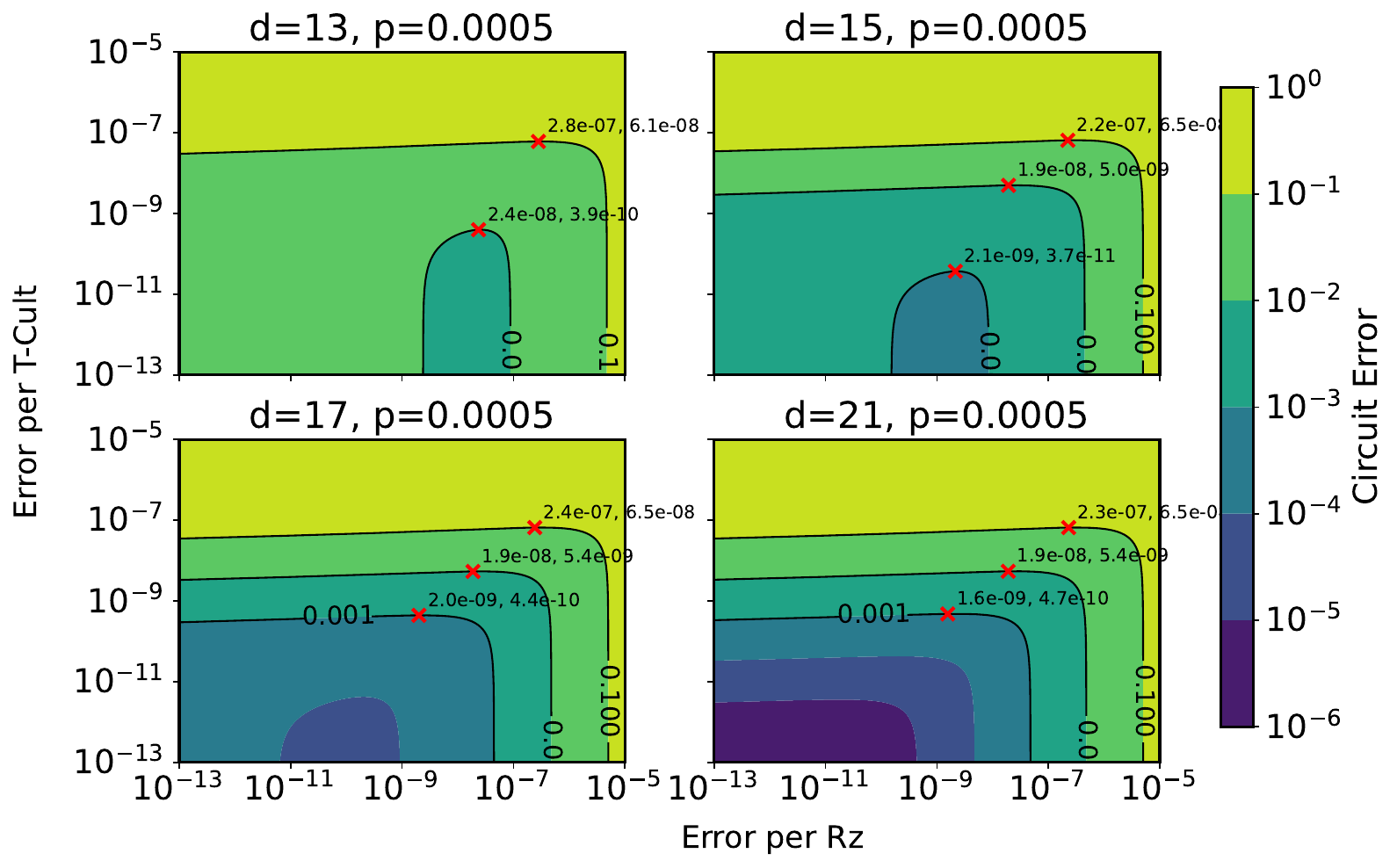} 
    \end{tabular}
    \caption{Sensitivity analysis of the relationship between code distance $d$, error per Rz $\epsilon_{R_Z}$, and error per cultivation $\epsilon_M$. The top row shows results for a 60 qubit Hamiltonian simulation circuit, while the bottom row shows results for a 100 qubit QAOA circuit. Each red `x' shows where the error per cultivate is maximized for each contour. In a cultivation-bottlenecked environment, a small change in the requested $|T\rangle$ state fidelity can have a substantial effect on the expected number of attempts (see Figure~14 of Ref.~\cite{gidney2024magicstatecultivationgrowing}). We also show the impact of two physical error rates: $p=0.001$ (Left) and $p=0.0005$ (Right). For lower error rates, small savings in $|T\rangle$ state fidelity are not as impactful, so we opt for the lowest available code distance.}
    \label{fig:sensitivity}
\end{figure*}
\begin{figure*}[t]
    \centering
    \includegraphics[width=\linewidth]{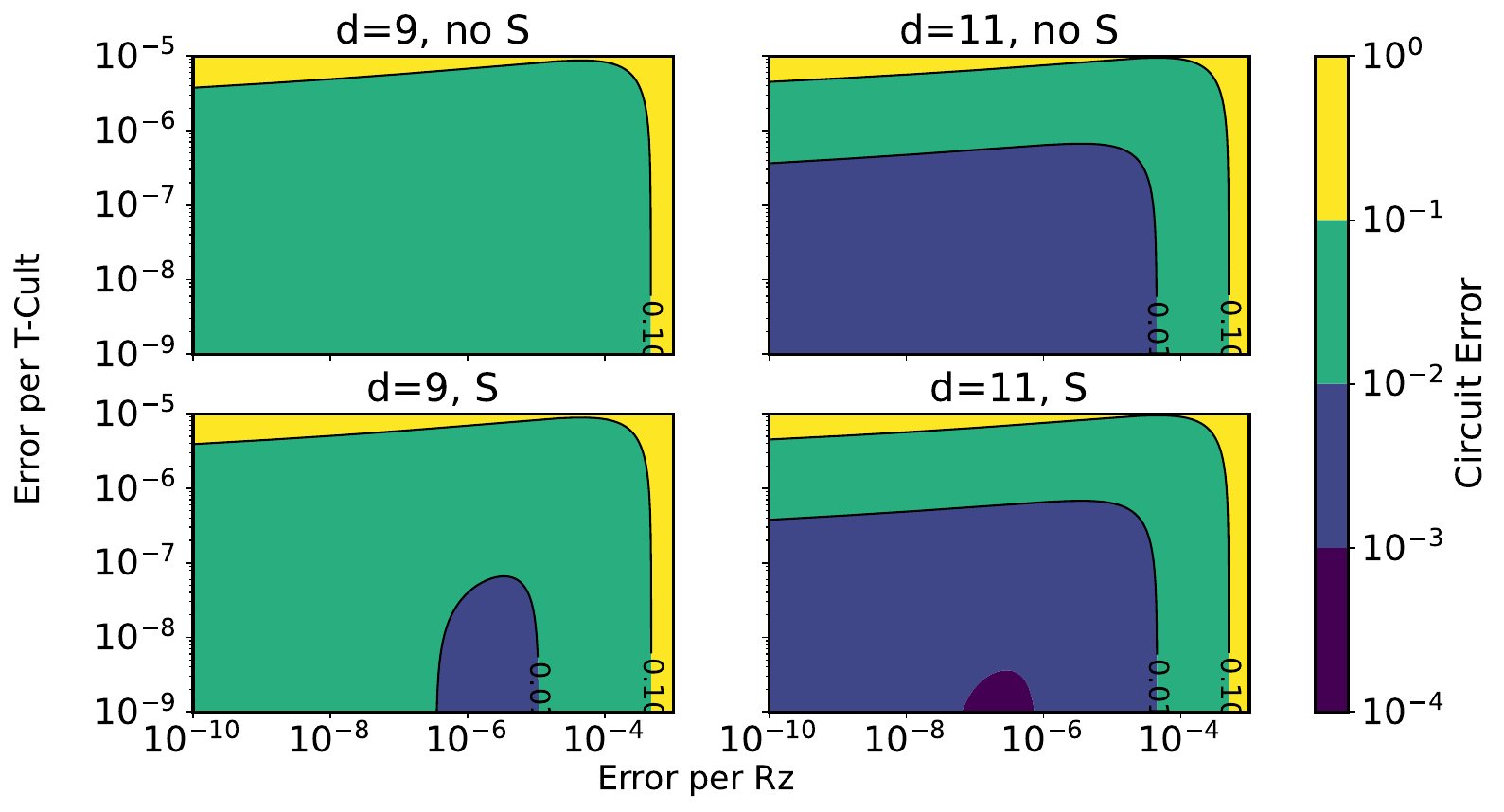}
    \caption{Sensitivity analysis plots assessing the relationship between code distance, cultivation error, Rz synthesis error, and overall circuit error between architectures with and without transversal S. These graphs use the same input Hamiltonian simulation circuit used in Section~\ref{sec:experiments} with the noise set to $p=0.0005$. The only difference in the calculation for the circuit error is the inclusion of the gates from teleportation, weighted according to needing to perform them only $50\%$ of the time.}
    \label{fig:s_trans}
\end{figure*}

\begin{figure*}[th]
    \centering
    \begin{subfigure}[t]{.6\linewidth}
        \centering
        \includegraphics[width=\linewidth]{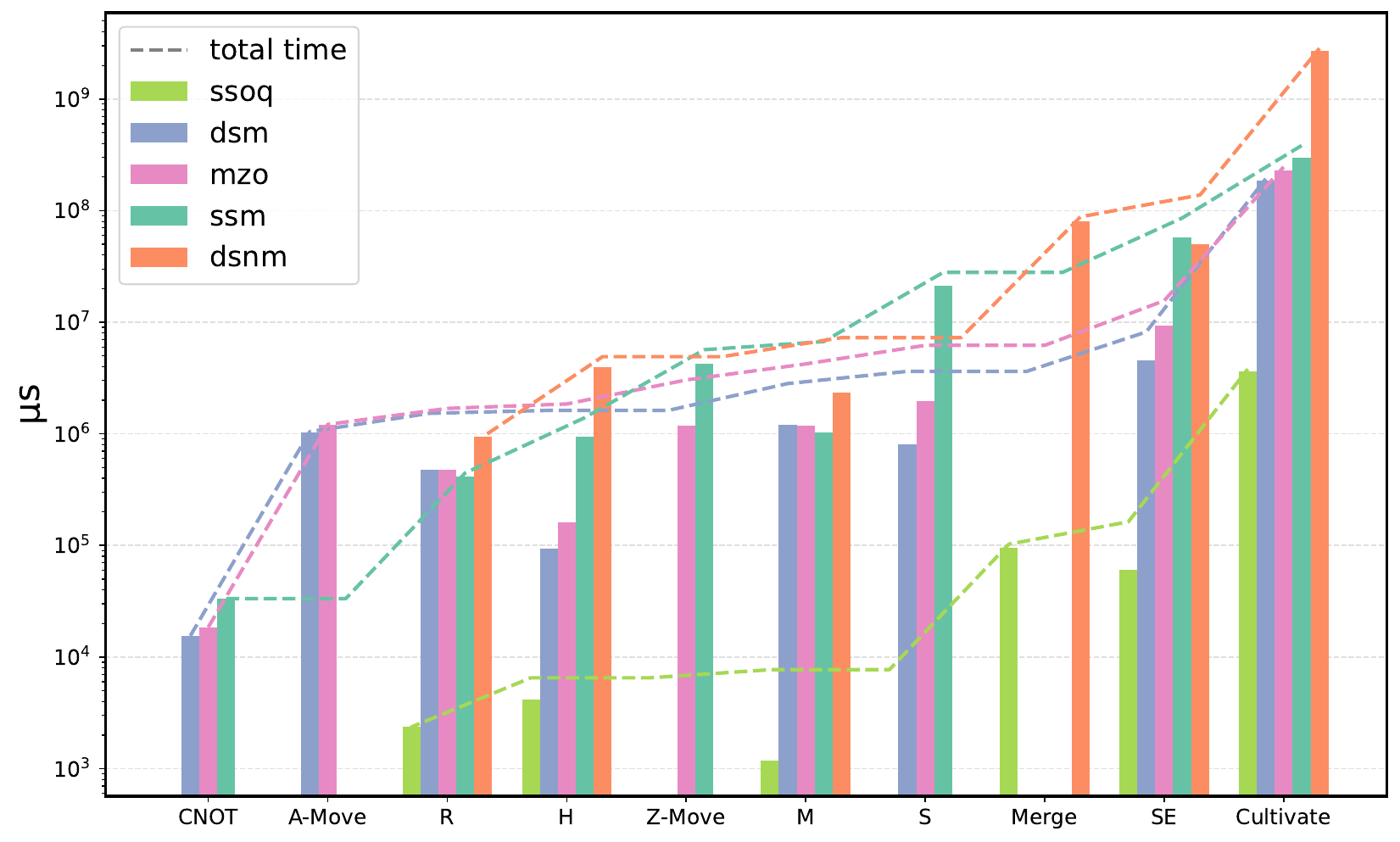}
        \caption{}
    \end{subfigure}\hfill
    \begin{subfigure}[t]{.4\linewidth}
        \centering
        \includegraphics[width=\linewidth]{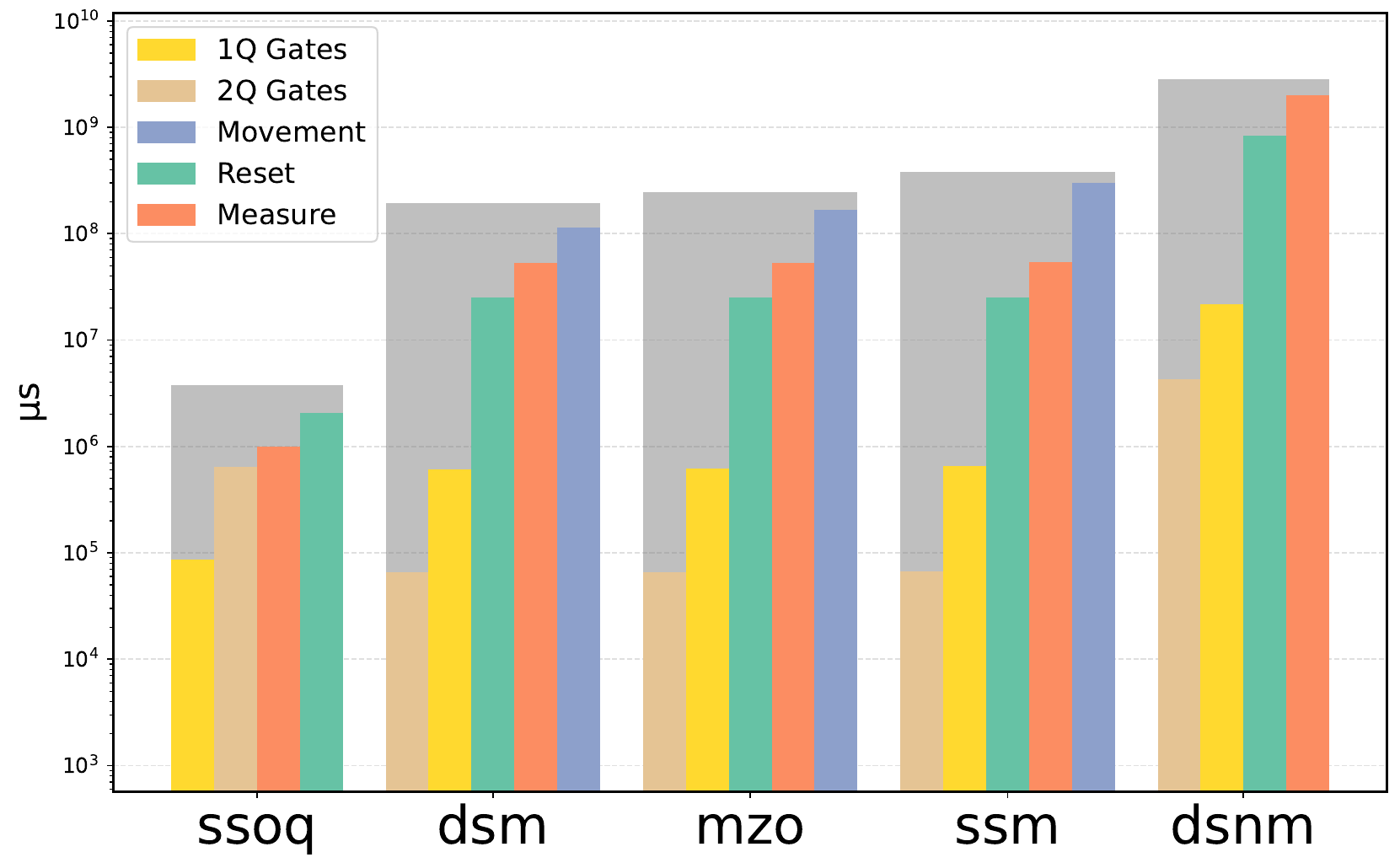}
        \caption{}
    \end{subfigure}
    \caption{Breakdown plots for a 60 qubit Hamiltonian simulation circuit with 10 Factory Qubits compiled with $99\%$ requested fidelity, using the ``Current'' columns of Table~\ref{tab:gate-speeds}. (a) Depicts cumulative time spent on each primitive in the critical path for each Architecture. The height of each bar represents the total time (in $\mu s$) contributed by the primitive on the x-axis to the final circuit time. The ``SE'' primitive includes contributions of one round of Syndrome Extraction for movement architectures and $d$ rounds of Syndrome Extraction for lattice surgery architectures. The dashed lines sit at the height of the current primitive added to all previous primitives to the left on the x-axis to emphasize the fact that contributions from the most expensive operations dwarf the combined total of all others. (b) Shows the total circuit time broken down by the physical gate operations one level below the primitives. The heights of the shaded regions correspond to the total circuit time of each Architecture. The contributions of A- and Z-Moves are combined under ``Movement''.
    }
    \label{fig:ham_sim}
\end{figure*}
\begin{figure*}[t]
    \centering
    \includegraphics[width=\linewidth]{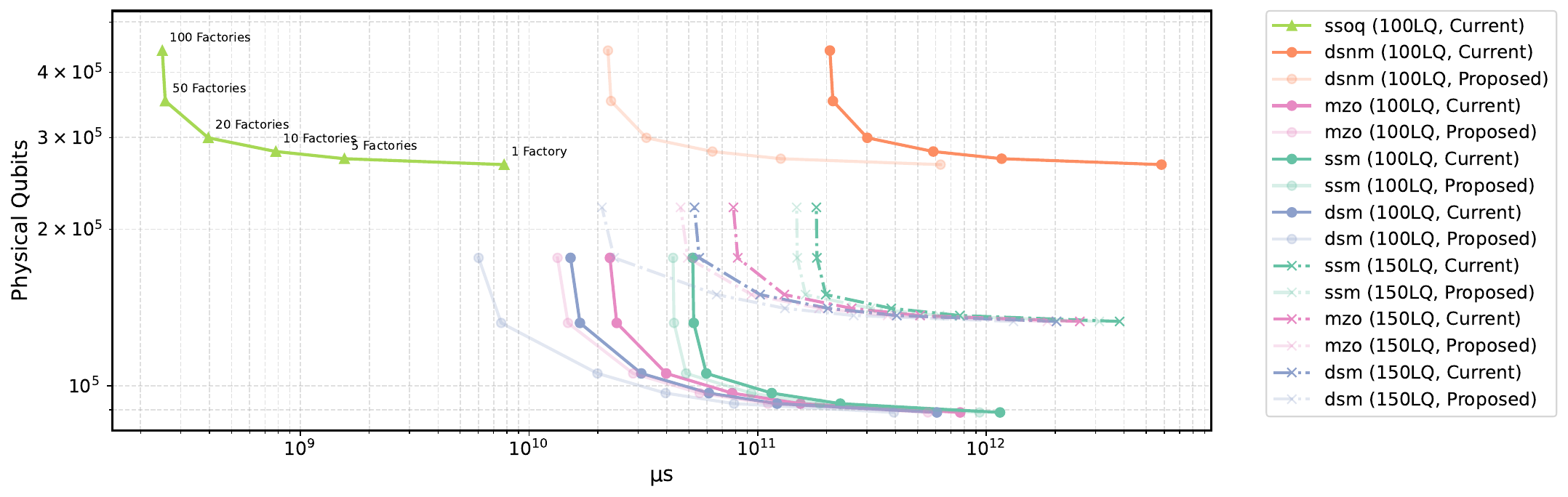}
    \caption{Factory comparison experiment for QAOA circuits at 100 and 150 logical qubits. The underlying problem and circuit reduction techniques used to produce the starting circuits come from Ref.~\cite{shah2025quantumenabledbiomarkerdiscoveryoutlook}. From right to left, each point on the curve corresponds to 1, 5, 10, 20, 50, and 100 factory qubits dedicated to $|T\rangle$ state cultivation. According to our model in Table~\ref{tab:arch-assumptions}, lattice surgery Architectures teleport $|S\rangle$ states, so they also include the overhead of the same number of S factories as T factories. Lower opacity lines represent the sped up versions of the Architectures using gate speeds in the ``Proposed'' column of Table~\ref{tab:gate-speeds}.}
    \label{fig:qaoa}
\end{figure*}
\begin{figure*}[t]
    \centering
    \includegraphics[width=\textwidth]{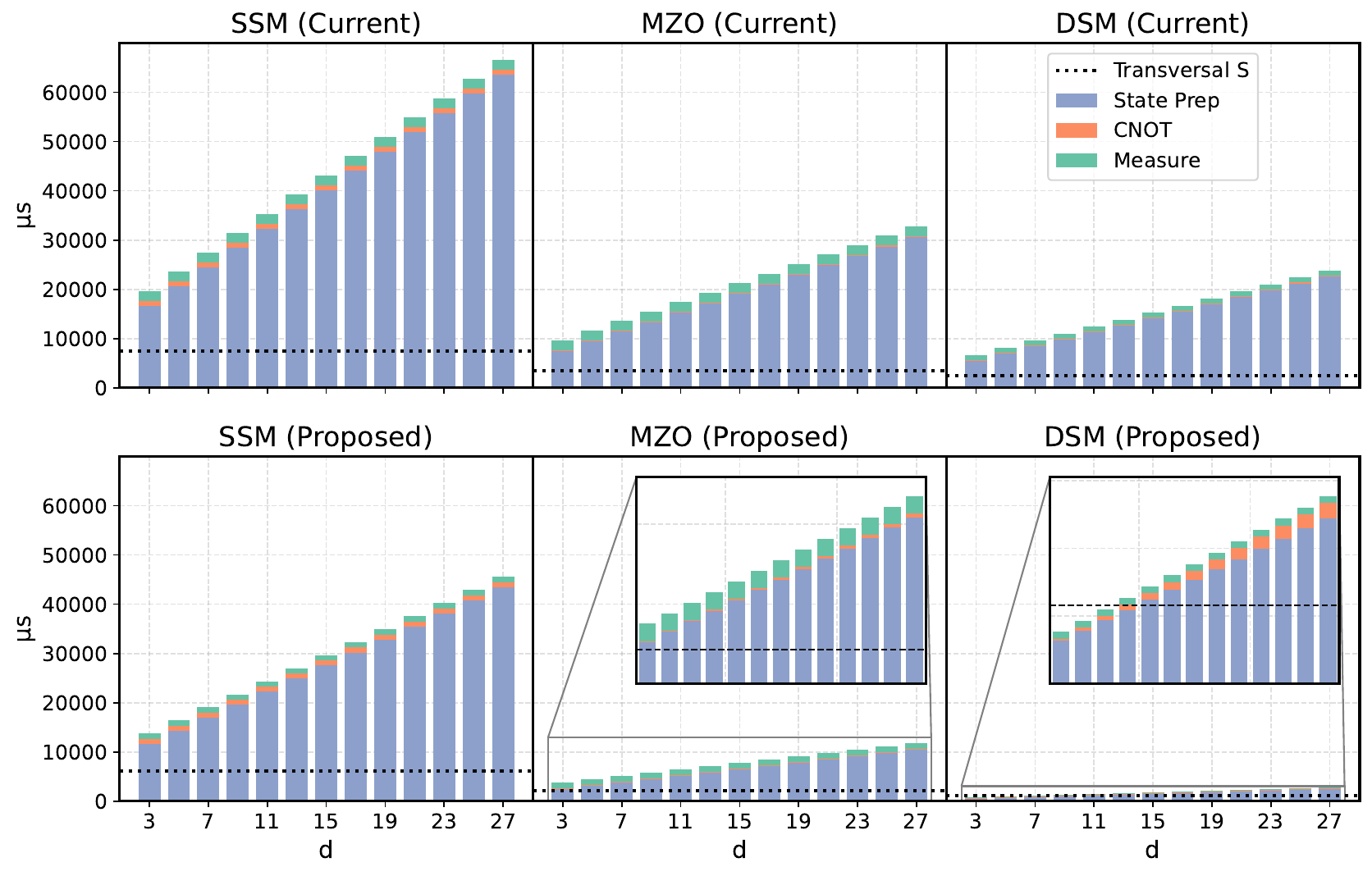}
    \caption{Comparison of teleporting S gates vs doing them transversally, broken down by component. State Prep, CNOT, and Measure all include time spent doing movement operations associated with readout and entanglement for the relevant architectures. Head-to-head times are comparable for only the fastest of architectures considered. Amortization through parallelism across Factory qubits can alter this relationship, but in an environment bottlenecked by magic state production, transversal S can benefit from being done while waiting for magic states.}
    \label{fig:s-inject}
\end{figure*}

\subsection{Error Sources}
\quad We now return to the question of choosing $\epsilon_{R_Z}$, as well as several other parameters based on error sources in the computation. First, let $k$ be the number of $R_Z$ gates in $C_1$, so that we can define the parameter $F_1:=\left(1-\epsilon_{R_Z}\right)^k$ to be the fidelity with which we approximate $C_1$ with $C_2$. $F_1$ can be thought of as the expected fidelity of the computation if all physical gates were perfect. However, the gates are not perfect, so we must account for the other sources of error. We subdivide the remaining error into two components. Let $F_2$ be the fidelity of operations in the QEC code and $F_3$ be the fidelity magic state production. Therefore the total fidelity of the circuit can be given by $F=F_1F_2F_3$.

To quantify $F_2$ we refer to Equation~\ref{eq1} as the relationship between hardware noise, code distance and fidelity. We use $\epsilon_L$ to denote the LER of normal operations within the code. Therefore if we know the number of Clifford operations in the Clifford + T circuit, then we can derive $F_2$. Similarly, we denote $\epsilon_M$ as the fidelity of the magic state cultivation procedure, and if we knew how many T gates were required in the Clifford + T circuit, we could derive $F_3$.

Unfortunately, the number of Clifford operations and T gates both depend on $\epsilon_{R_Z}$. Requiring greater precision in the approximation of single qubit gates also incurs a greater cost of Cliffords and T gates to approximate them. Figure~\ref{fig:synth_plot} shows the relationship between the requested precision in the single qubit gate and the number of each type of gate. We empirically determine that the synthesis adds approximately $5\log\epsilon_{R_Z}$ T gates and $8\log\epsilon_{R_Z}$ Clifford gates. Using these numbers, we can give an equation for the total circuit fidelity as a function of the error per $R_Z$ gate, code LER, magic state LER, and gate counts $k,l$.
\begin{align*}
&F(\epsilon_{R_Z}, \epsilon_M,\epsilon_L) = \\
&\left(1-\epsilon_{R_Z}\right)^k
\left(1-\epsilon_M\right)^{5k\left|\log\epsilon_{R_Z}\right|}
\left(1-\epsilon_L\right)^{l+8|\log\epsilon_{R_Z}|}
\label{eq:circuit-fid}
\end{align*}
The function $F$ can be further broken down using the expression in Equation~\ref{eq1} to replace the dependence on $\epsilon_L$ with a dependence on code distance $d$ (for fixed hardware noise).

We propose two methods to choose these parameters. This first and easiest is to divide the total circuit error budget in half, allotting one half of the total error to synthesizing $R_Z$ gates and the other half to logical operation error. Determining $\epsilon_{R_Z}$ can be done directly from the now separate error budget, and the extra gates from synthesis become fixed. Since the cultivation fidelity is typically much lower than the code fidelity, we optimize for that next. For our implementation, we can produce either a weak cultivation error of $10^{-6}$ or a strong cultivation error of $10^{-9}$. If weak cultivation is enough, we choose it. Otherwise we choose $10^{-9}$. Then we choose the minimum code distance such that the total circuit error remains under the tolerance.

Another method for optimizing these parameters is to perform a sensitivity analysis of $F$ to find the precise combination of numbers that gives the best outcome\footnote{The equation optimized in practice includes extra errors from teleportation sub-circuits and corrections, which themselves may also require teleportation.}. Figure~\ref{fig:sensitivity} shows an example sensitivity analysis for two of the circuits in Section~\ref{sec:experiments}.

Typically, the first method is sufficient to perform meaningful architectural comparisons, especially for larger, deeper circuits. Nevertheless, in early fault tolerant applications, the second method can yield fruitful results. For instance, we found that using our model on the Hamiltonian circuit in Section~\ref{sec:experiments}, $99\%$ fidelity can be achieved with a lower code distance for an architecture with access to transversal S than without it (Figure~\ref{fig:s_trans}).

\subsection{Architecture Instantiation}
\label{subsec:architecture-and-layout}
\quad The next step in the pipeline is initializing an Architecture, choosing a Layout, and setting the keyword arguments. Together, these decisions determine the most important assumptions about the hardware expected to run the circuit in the early fault tolerant regime.

The most important aspect of the Architecture is determining its access to movement. This decision impacts the available primitives, the options for Layout configuration, and access to folded cultivation and correlated decoding (two of the keyword arguments). Movement Architectures are further differentiated based on when movement is necessary, while the lattice surgery architectures considered here only differ in their gate speeds. At the time of writing, our software supports the Architectures listed in Table~\ref{tab:arch-assumptions}. Additionally, all Architectures use the gate speeds in Table~\ref{tab:gate-speeds} as default, but gate speeds can be altered easily to experiment with faster or slower gates.

The previous step in the pipeline can be used to set some of the keyword arguments. The code distance $d$ can come directly from the previous step, and the cultivation repetition parameter can be derived from the magic state fidelity $\epsilon_M$. Whether determined through sensitivity analysis or through choosing between the weak or strong cultivation fidelities, the cultivation repetition parameter follows as a direct consequence. Under the standard assumption of $p=0.001$ and using the cultivation procedure in Ref.~\cite{gidney2024magicstatecultivationgrowing}, the end-to-end retry rate for cultivation is $80\%$ for an error of $10^{-6}$ and $99\%$ for an error of $10^{-9}$, which lead cultivation repetition parameters of $5$ and $100$, respectively. Using the folded version of the procedure typically yields a repetition parameter approximately $10$ times smaller. Though closely related to the magic state fidelity, the cultivation repetition parameter remains a keyword argument to allow for experimentation with different procedures that share the repeat until success character of cultivation.

The choice of Layout provides an opportunity to control the number of T and S factories. More factories will decrease the average time expected to wait for a $\ket{T}$ state at the cost of more physical qubits. This effect is more prevalent in the lattice surgery Architectures where whole ancilla code patches are dedicated to facilitating entanglement between distant qubits. On the other hand, as described in Section~\ref{sec:core-components}, our Layout strategy for movement Architectures is simple.

To consider the impact of different Layouts for lattice surgery architectures, we implement several different generation methods---two based on access to a fixed number of factories and a qubit-frugal one where the number of T and S factories can be chosen independently. We found that the qubit frugal option performed the best both in terms of space and time for the circuits considered. Appendix~\ref{app:layout-generation} contains more detail on the Layouts considered. Choosing an optimal logical layout is hard in general, and many application specific resource estimates go to great lengths to determine good layouts that work for their specific applications. However, to experiment with a broader class of circuits, we let our Layouts be flexible, while having several useful starting points that can be generated procedurally.

The initialization of the Architecture completes the necessary groundwork for the proceeding compilation to primitives. It serves as a valuable framework that makes clear which assumptions about the underlying hardware are included in the model, and it creates a unique opportunity to rapidly iterate in response to changing ideas about the device.

\subsection{Primitive Compilation}
\label{subsec:primitive-compilation}
\quad With the Architecture and Layout configured, we implement a simple ``read-and-replace'' compiler. First we decompose the circuit into primitives. For movement-based architectures, all of the Clifford operations can be done transversally. For lattice architectures, the Clifford operations are more complicated. CNOT operations are routed and decomposed into Merge and Split primitives. While not truly transversal, H does not have to be routed because lattice Layouts always have an available ancilla patch that can be used for the physical decomposition of the H primitive. S gates are treated similarly to T gates. The main work of the compiler is to coordinate the production of $\ket{T}$ states (or $\ket{S}$ states for lattice architectures). Both $\ket{S}$ and $\ket{T}$ state generation takes place on a factory qubit by applying the relevant Cultivation operation. The compiler greedily chooses the closest available factory to use as fuel for gate teleportation. The logical circuits used to teleport T with and without a transversal S are shown in Figure~\ref{fig:teleportation-circuits}. After the factory has been used, the Cultivation operation is applied again, so that the factory can be used again over the course of the circuit.

We note several simplifying assumptions made for the compilation procedure. The outcome of the measurement operations in gate teleportation is random. Therefore, applying the S correction after the first CNOT in figure \ref{fig:teleportation-circuits} will not be necessary approximately half the time; however, to keep estimates conservative and results deterministic, our compiler assumes the corrections are always needed when teleporting gates. Another simplification is the presumption that $\ket{T}$ ($\ket{S}$) states have been successfully cultivated immediately after the Cultivation operation. In practice, running the circuit would involve dynamically choosing the closest available factory based on whether it has successfully produced a T state or not, which is a non-deterministic process. To capture some of this behavior, we have the compiler treat used factories as being ``unavailable'' until all factory qubits have been used, at which point the Cultivation operation is called on all factory qubits at once. This choice does not change the logic of the circuit, but it does create the opportunity for some logical qubits to require factories at greater distances than they would otherwise use if immediate Cultivation success were assumed.

Next the compiler performs passes for post-op-correction and idling. If the post-op-correction flag is engaged, the compiler adds syndrome extraction after each transversal operation in the circuit with the number of rounds of extraction equal to the number of rounds requested by the architecture, allowing for experimentation with looser syndrome extraction requirements like correlated decoding. Similarly, the idling pass adds Syndrome Extraction to logical qubits without operations in each moment. While this pass does not impact the total circuit duration, it can be useful to consider the impact of idling on the total gate count. Lastly, if the architecture has access to movement, the compiler performs a pass adding logical movement operations according to the architectures movement type, differentiating between movements to a zone and movements within the logical qubit array.

\subsection{Resource Estimates}
\label{subsec:resource-estimates}

\quad Having the circuit compiled to Primitive operations makes generating resource estimates relatively simple. If the compiled circuit has $n$ logical qubits (including factories), the number of physical qubits is $d^2n$. The total number and kind of each physical operation is obtained by adding up the resources of the pre-compiled physical circuits for each primitive, weighted by the number of occurrences in the Primitive circuit. To determine the total circuit time, we find its critical path. The critical path is the longest sequence of operations that must be done sequentially to complete the circuit. To find the critical path we use the times for each primitive saved in the Architecture. Since the critical path calculation takes place at the level of primitives, it assumes full parallelism in the underlying operations. One way this assumption can potentially break down is the fact that it is likely unreasonable to expect that a logical reset operation, which might involve atom reloading or optical pumping, will be able to take place at the same time as an entire Cultivation operation. However, the relative expense of magic state production allows us to be more comfortable with a coarser picture with the assumed high degree of parallelism. We also track serial cost of the operations to serve as an upper bound on the circuit time without any parallelism. Determining the right mix of parallelism at the level of primitives is a question for future work.

With the large number of configurable parameters between the Architecture and Layout and the freedom of an arbitrary circuit as input, these resources can easily be recomputed according to a different set of assumptions, often without having to re-run the compiler. The primitive circuit makes clear the impact of different gate speeds on the final resources by recalculating the critical path with updated operation times. While large reconfigurations---like access to correlated decoding or restricting the movement type---impact the Primitive circuit, many questions can be posed without changing the intermediate representation.

A further benefit of the fault tolerant compilation is the ability to break down the cost of a circuit according to either its physical operations or its primitive operations. Collecting the primitive operations in the critical path and knowing their physical decompositions and operation times from the architecture allows us to assess resources in greater detail, as in Figure~\ref{fig:ham_sim}. The level of abstraction in this pipeline is still well above the physical level description of the operations, but it is much more closely aligned with the main components of interest to teams building early fault tolerant quantum computers based on neutral atoms.

\section{Experiments}
\label{sec:experiments}
\quad We now show how to use the tool to perform some useful experiments on different neutral atom architectures. We reconfigure gate speeds, consider access to different types of movement, assess spacetime tradeoffs, and investigate sensitivity. We focus on two candidate circuits for early fault tolerance: trotterized Hamiltonian simulation and quantum optimization. These problems fall within the regime of what might be possible with neutral atoms without connecting modules via photonic interconnects~\cite{PRXQuantum.6.010101}. They also have the character of varying significantly between inputs and using many continuous angle rotations, unlike arithmetic logic circuits.
\subsection{Circuit Experiments}
\quad To compare how different architectures might perform on Hamiltonian simulation, we use a 60 logical qubit instance of the Kanamori Hamiltonian~\cite{kanamori1963electron, georges2013strong}. The circuit is motivated by the material $\text{SrVO}_3$, which can function as an important
benchmark for strongly correlated materials. See Appendix~\ref{app:kanamori} for more details on the underlying application.

Figure~\ref{fig:ham_sim} shows the primitive breakdown for a single Trotter step of the Hamiltonian simulation circuit with a target circuit fidelity of $99\%$ across the architectures in Table~\ref{tab:arch-summary}. The simple pipeline initially led us to choose $d=13$, $\epsilon_{R_Z}=2.4\times 10^{-5}$, and $\epsilon_{M}=10^{-9}$; however, using the sensitivity analysis in Figure~\ref{fig:sensitivity}, we were able to get slightly more favorable parameters. We found that $d=15$, $\epsilon_{R_Z}=2.7\times 10^{-6}$, and $\epsilon_M=6.1\times10^{-7}$ offered a reduction in circuit time at the cost of more physical qubits. Increasing the code distance allows for a greater tolerance in the the magic state error, reducing the cultivation repetition parameter from $100$ to $40$ using basic cultivation and a similar reduction for folded cultivation. Since cultivation is the main bottleneck for this application, we chose parameters to reduce its impact as much as possible. This experiment highlights the power of our model to consider different input circuits much more rapidly to understand the impacts of architectural decisions for different applications without having to spend months hand-crafting a new analysis. For this application, we observe that Cultivation comprises over $90\%$ of the total circuit time for all architectures.

Our approach offers the unique opportunity to measure the circuit resources at multiple layers of abstraction. By compiling to the primitive circuit, we are able to see which primitive operations are the most costly to total circuit time. We can also break down that same information into the underlying physical gates.

We also considered a large, fault tolerant QAOA circuit at depth $1$ at $\geq 100$ qubits. The results are shown in Figure~\ref{fig:qaoa}. We demonstrate the impact of adding more factory qubits to trade physical qubit space for shorter run time. For each analysis, increasing the number of factories produced over a $10$x speedup until reaching a point of diminishing returns. For this example, this point was approximately 10 factories.

Additionally, we analyze the impact of increased gate speeds according to the proposed improvements to Measurement and Reset in Table~\ref{tab:gate-speeds} to make an informed (but still conservative) estimate into what might be possible in the future. We find that there is a regime where neutral atom and superconducting architectures are competitive, despite the substantial difference in gate speeds. To get there, our model suggests that neutral atoms might need access to every advantage we consider: correlated decoding, transversal gates, frugal movement, folded cultivation (which requires three-qubit gates), and $10$x faster measurements than the current state of the art. A more detailed model for movement, improved compilation techniques at different levels of the stack, and a careful choice of input quantum circuit could potentially offer opportunities to close this gap. Rapidly determining where different architectures are competitive over many configurations and applications is where our approach excels.
\subsection{The Benefits of Transversal S}
\label{subsec:cult-vs-teleport}
\quad In addition to the overall circuit resources, it is possible to use the tool to consider alternate implementations of primitive operations. Previous work~\cite{sunami2025} chooses not to use the fold transversal S gate available to neutral atoms, arguing that it would be faster to cultivate them instead. This argument mainly relies on the potential to amortize the cost over several factories, allowing the time to approach zero. In Figure~\ref{fig:s-inject}, we consider a wide range of scenarios for comparing transversal S to gate teleportation from a resource state. Using the ``Stationary S Gate'' described in Ref.~\cite{Gidney2024inplaceaccessto} as the cost of cultivation, we find that without parallelization, teleportation is nearly universally more expensive. The only regime where the two are competitive is for low code distances and the fastest considered architectures. 

Of course, the parallelization factor can make a difference. Ignoring briefly the space cost of many patches producing $\ket{S}$ states, one can imagine the Cult-S primitive to have a negligible time; However, the remaining serial component is the teleportation circuit. This component requires a logical entangling gate accompanied by a movement and a logical measurement. Using the primitive times in Table~\ref{tab:primitive-times} it is clear that the time required for some architectures to perform CNOT, Move, and Measure relative to transversal S is not trivial. For the SSM architecture, replacing the transversal S gate with CNOT, Move, and Measure takes the operation from $6156\mu s$ to $1510\mu s$, an improvement of about $75\%$, but for the DSM architecture, the improvement is only about $100\mu s$, or about $10\%$ The benefits seem especially meager considering DSM $d=11$ is the regime in Figure~\ref{fig:s-inject} where the amortization is needed the least. To sacrifice the qubit space that could otherwise go toward reducing the cost of $\ket{T}$ state cultivation appears to be a better choice for early fault tolerant applications.

One more reason to consider choosing transversal S over gate teleportation is to save the extra gate cost involved in teleportation. This cost is small for large code distances where errors are suppressed significantly, but when experimenting with low-noise systems performing sensitivity analysis for the Kanamori circuit, we found that access to transversal S was the sole reason for needing a higher code distance, and consequently quadratically more qubits for the circuit to remain under $1\%$ circuit error. The results are shown in Figure~\ref{fig:s_trans}.

\begin{figure*}[ht]
    \centering
    \includegraphics[width=\textwidth]{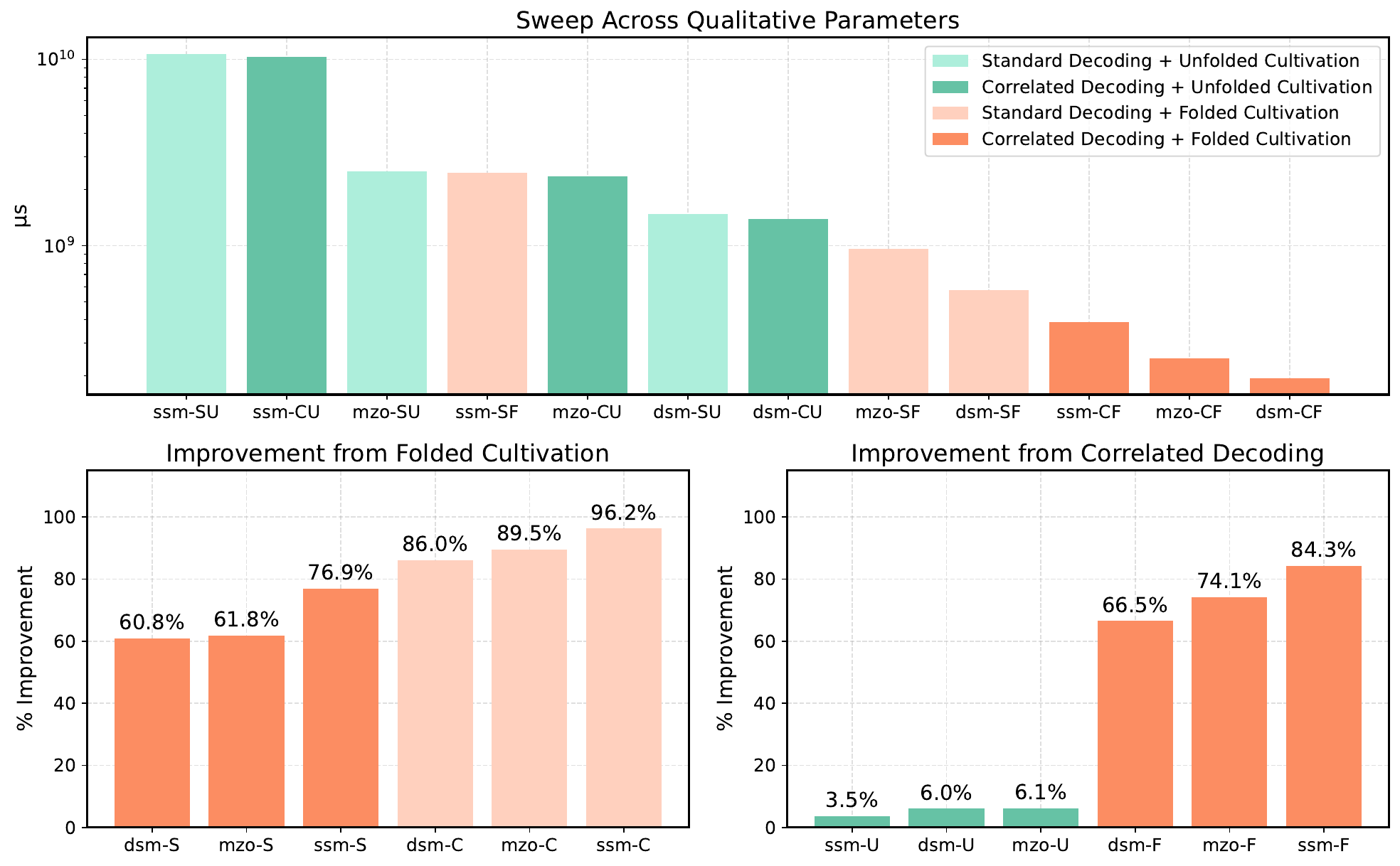}
    \caption{Parameter sweep across different Architectures, cultivation styles, and decoding strategies for the Hamiltonian simulation circuit in Section~\ref{sec:experiments}. Labels use the same Architecture acronyms from Table~\ref{tab:arch-summary} and current gate speeds according to Table~\ref{tab:gate-speeds}. Two additional labels are included: C/S denotes correlated (C) vs standard (S) decoding, and U/F denotes unfolded (U) vs folded (F) cultivation styles. Across the board, folded cultivation improves circuit time by at least 50\%, but has an even higher performance improvement factor on Architectures using correlated decoding. Conversely, correlated decoding only shows a significant effect on architectures already using folded cultivation.}
    \label{fig:param-sweep}
\end{figure*}
\subsection{Parameter Sweep}
\quad With many parameters at the Architecture's disposal, a key question in assessing the quantum resources is which parameters are most important. Our framework makes it possible to answer these questions quantitatively for specific circuits. In Figure~\ref{fig:param-sweep} we examine the joint impacts of fold transversal cultivation and correlated decoding on the runtime of the Hamiltonian simulation circuit. In the previous analyses of this circuit in this work, we used both to demonstrate the best performance; however, looking at the effects of these two qualitative parameters individually can be illuminating on their relative importance to improving the overall circuit time.

While having both is best, we find that the transversal cultivation speedup is consistently more effective, offering an increase in speed of execution of at least 60\% for each Architecture instance. On the other hand, correlated decoding only seems to offer a similar improvement factor when stacked on top of the benefits of folded cultivation.

With only a space of two yes/no questions on top of our three baseline movement-based architectures the space of possible combinations starts to become large. The ability to consider each of these combinations on a per-circuit basis is valuable for finding Architectures that fit well with applications.

\section{Outlook and Conclusion}
\label{sec:conclusion}
\quad In this work we have approached the problem of quantum resource estimation for early fault tolerant quantum computers as a compilation problem, with an emphasis on solving this problem for neutral atom architectures. We build upon the heavy circuit optimization strategies of NISQ, to create a carefully constructed set of building blocks called primitives and provide their decompositions to neutral atom gatesets. We use this set of operations, information about the logical qubit layout, and keyword arguments to construct the Architecture class, which we use as the base of our compilation stack.

Utilizing the broad work on the rotated surface code, we walk through the complete process of resource estimation from the input circuit to the key resources. We also demonstrate how our tool enables us to make rapid architectural comparisons over a large set of assumptions for early fault tolerant applications to derive key insights. Neutral atoms are expected to grow to larger systems before needing networking between QPUs, making the low encoding rate of the rotated surface seem less limiting. Our work highlights the need for improvement from both hardware and theory to run early fault tolerant circuits. In particular, we identify magic state cultivation as the primary bottleneck for these systems in terms of runtime, while keeping open the possibility of spacetime tradeoffs.

Our work enables specific architectural comparisons based on clearly defined numerical parameters that are themselves verifiable in the literature. Furthermore, we make our model flexible to incorporate changes in these assumptions, allowing our work to grow with the evolution of the field. We showcase the current capability of the tool by performing a series of experiments that would be difficult to perform without our work.

Our analysis highlights the importance of access to movement as a key method of saving quantum resources. We conclude that relying on lattice surgery alone in neutral atom systems will not be sufficient, even for early fault tolerant applications. At the same time, our cross-architectural analysis gives insight into the importance of being frugal with movement. The results presented in this work demonstrate favorable advantages for neutral atom approaches combining dual species and movement. Despite movement's ability to unlock key features for an architecture, it quickly becomes the main bottleneck for running real circuits. Refining our model of atom movement by incorporating considerations for atom re-cooling and standardized moves lays out a clear path for improvements to our tool.

As neutral atom quantum computers continue to develop both on the hardware and theoretical side, we expect our work will be a natural way to include more hardware-focused resource estimation experiments to evaluate different fault tolerant architectures. We look forward to future experiments on real devices that will help refine our projections for cultivation times. These experiments will likely yield hardware-focused optimizations in turn that will allow us to improve our model. The potential for this iterative process highlights the value of heavy circuit optimization characteristic of the NISQ era and extends its importance to saving resources for circuits in the early fault tolerant era.

Our work has yielded several promising avenues for future work. The prevalence of continuous angle rotations in many mid-size quantum applications motivates more consideration for other methods for single qubit gate synthesis and more broadly available tools for performing general Clifford + T compilation. There is also great value in building upon our framework to expand our analysis to comparison between more QEC codes on top of our architectural comparisons. This work primarily focuses on the surface code, but comparison with other codes may yield new insights and innovations in this space. There remain many important considerations to include in future models that will improve our understanding of early fault tolerant applications that we expect will fold well into our compiler framework.

\section*{Data Availability}
\quad The code and data for this research can be found at \url{https://github.com/Infleqtion/resource-superstaq}.

\section*{Acknowledgements}
\label{sec:acknowlegements}
\quad This research is supported by the U.S. DOE Office of Science-Advanced Scientific Computing Research Program, under Contract No. DE-AC02-06CH11357. The funder played no role in study design, data collection, analysis and interpretation of data, or the writing of this manuscript.
This work is supported in part by Wellcome Leap as part of the ‘Quantum Biomarker Algorithms for Multimodal Cancer Data’ research project within the Quantum for Bio (Q4Bio) Program.

The authors thank Joshua Viszlai for numerous helpful conversations about quantum error correction and magic state cultivation.

\def\bibsection{\section{References}}
\bibliographystyle{unsrt}
\bibliography{refs}

\onecolumn
\appendix
\onecolumn

\begin{table}[p]
\centering
\begin{subtable}{\textwidth}
\centering

\setlength{\tabcolsep}{4pt}      
\renewcommand{\arraystretch}{1.15} 

\resizebox{\linewidth}{!}{%
\begin{tabular}{cccccc}
\hline
\textbf{\shortstack{Architecture \\ Name \\ (Acronym)}} &
\textbf{\shortstack{Qubit\\Modality}} &
\textbf{\shortstack{Gate\\Speeds\\(See Tab.~\ref{tab:gate-speeds})}} &
\textbf{\shortstack{Qubit\\Movement}} &
\textbf{\shortstack{In-place\\Entanglement}} &
\textbf{\shortstack{In-place\\Readout}} \\
\hline
\shortstack{Single Species\\with Movement\\(SSM)} & Rb atoms & Neutral atom & \cmark & \xmark & \xmark \\
\hline
\shortstack{Measurement Zones\\Only (MZO)} & \shortstack{Cs atoms} & Neutral atom & \cmark & \cmark & \xmark \\
\hline
\shortstack{Dual Species\\with Movement\\(DSM)} & Cs and Rb atoms & Neutral atom & \cmark & \cmark & \cmark \\
\hline
\shortstack{Dual Species\\No Movement\\(DSNM)} & Cs and Rb atoms & Neutral atom & \xmark & \cmark & \cmark \\
\hline
\shortstack{Superconducting\\Sea-of-Qubits\\(SSOQ)} & Transmons & Superconducting & \xmark & \cmark & \cmark \\
\hline
\end{tabular}%
}
\caption{Hardware Assumptions for several neutral atom approaches to fault tolerant quantum computing and one representative of superconducting transmon qubits. SSM is based on zoned architectures like Ref~\cite{harvard-arch}. MZO takes SSM and adds in-place entanglement based on Ref.~\cite{rines2025demonstration} DSM further adds in-place readout/entanglement enabled by dual species~\cite{fang2025interleaved}. Lastly, DSNM and SSOQ represent fixed qubit architectures with nearest-neighbor connectivity restrictions and support differing gate speeds given by Table~\ref{tab:gate-speeds}.}
\label{tab:hw-assumptions}
\end{subtable}

\vspace{1.8em}

\begin{subtable}{\textwidth}
\centering

\setlength{\tabcolsep}{6pt}
\renewcommand{\arraystretch}{1.15}

\resizebox{\linewidth}{!}{%
\begin{tabular}{ccccc}
\hline
\textbf{Architecture} &
\textbf{\shortstack{Logical\\Entanglement}} &
\textbf{\shortstack{T state\\Preparation}} &
\textbf{\shortstack{S gate\\Implementation}} &
\textbf{\shortstack{Syndrome\\Rounds}} \\
\hline
SSM & Transversal gates & \cite{gidney2024magicstatecultivationgrowing} & \cite{chen2024transversallogicalcliffordgates} & $1$ \\
\hline
SSM (Fold) & Transversal gates & \cite{sahay2025foldtransversalsurfacecodecultivation} & \cite{chen2024transversallogicalcliffordgates} & $1$ \\
\hline
MZO & Transversal gates & \cite{gidney2024magicstatecultivationgrowing} & \cite{chen2024transversallogicalcliffordgates} & $1$ \\
\hline
MZO (Fold) & Transversal gates & \cite{sahay2025foldtransversalsurfacecodecultivation} & \cite{chen2024transversallogicalcliffordgates} & $1$ \\
\hline
DSM & Transversal gates & \cite{gidney2024magicstatecultivationgrowing} & \cite{chen2024transversallogicalcliffordgates} & $1$ \\
\hline
DSM (Fold) & Transversal gates & \cite{sahay2025foldtransversalsurfacecodecultivation} & \cite{chen2024transversallogicalcliffordgates} & $1$ \\
\hline
DSNM & Lattice surgery & \cite{gidney2024magicstatecultivationgrowing} & \cite{Gidney2024inplaceaccessto} + Teleport & $d$ \\
\hline
SSOQ & Lattice surgery & \cite{gidney2024magicstatecultivationgrowing} & \cite{Gidney2024inplaceaccessto} + Teleport & $d$ \\
\hline
\end{tabular}%
}
\caption{Architectural Assumptions for key primitives. Ref.~\cite{gidney2024magicstatecultivationgrowing} contains details for magic state cultivation in a nearest-neighbor setting. Ref~\cite{sahay2025foldtransversalsurfacecodecultivation} extends the protocol to architectures with higher connectivity and access to neutral atom gates. Ref.~\cite{Gidney2024inplaceaccessto} shows how to cheaply prepare $|S\rangle$ states. Ref.~\cite{chen2024transversallogicalcliffordgates} shows how to apply transversal S with access to movement. Syndrome Rounds refers to the number of rounds of syndrome measurement, depending on if correlated decoding is an option.}
\label{tab:arch-assumptions}
\end{subtable}
\centering
\caption{Summary of the assumptions made for all of the architectures considered in this work.}
\label{tab:arch-summary}
\end{table}

\begin{table}[tbp]
\centering
\begin{tabular}{l|cc|c}
& \multicolumn{2}{c|}{\textbf{Neutral atoms}} & \textbf{Superconducting} \\
& Current \cite{harvard-arch, Zhou_2025} & Proposed \cite{scott2025laser} & Current \cite{Barends_2014, bengtsson2024model} \\ \hline
1Q Gates       & 5     & -   & 0.02 \\
2Q Gates       & 0.27  & -   & 0.04 \\
Measure        & 1000  & 100 & 0.5 \\
Reset          & 400   & 40  & 1 \\
Z-Move & 500   & -   & -    \\
A-Move & 20-500 & - & - \\
\end{tabular}
\caption{Physical operation times for neutral atom and superconducting modalities. Movement is broken into two types, A-Move and Z-Move. A-Move is a movement between lattice site alleys to align data qubits for in-place entanglement, while Z-Move is a movement to an entangling or readout zone. All times are in $\mu s$. The Reset operation in the Superconducting column represents having a ``cycle time'' of $1\mu s$.}
\label{tab:gate-speeds}
\end{table}


\begin{table*}[t]
\centering
\resizebox{1\textwidth}{!}{
\begin{tabular}{c|cccccccc}
  \hline
  \textbf{Current} & \textbf{SSM} & \textbf{SSM (Fold)} & \textbf{MZO} & \textbf{MZO (Fold)} & \textbf{DSM}  & \textbf{DSM (Fold)} & \textbf{DSNM} & \textbf{SSOQ} \\ \hline
  I & 0 & 0 & 0 & 0 & 0 & 0 & 0 & 0 \\
  Z & 0 & 0 & 0 & 0 & 0 & 0 & 0 & 0 \\
  X & 0 & 0 & 0 & 0 & 0 & 0 & 0 & 0 \\
  H & 505 & 505 & 505 & 505 & 505 & 505 & 33,849 & 40\\
  S & 7,416 & 7,416 & 2,411 & 2,411 & 1,411 & 1,411 & - & -\\
  CNOT & 10 & 10 & 10 & 10 & 10 & 10 & - & - \\
  Measure & 1,000 & 1,000 & 1,000 & 1,000 & 1,000 & 1,000 & 1,000 & 1 \\
  Reset & 400 & 400 & 400 & 400 & 400 & 400 & 400 & 1 \\
  SE & 6,411 & 6,411 & 2,411 & 2,411 & 1,411 & 1,411 & 1,411 & 19 \\
  Cult-S & 32,319 & 32,319 & 15,319 & 15,319 & 11,319 & 11,319 & 11,319 & 14 \\
  Cult-T & 17,201,591 & 719,277 & 3,901,591 & 484,277 & 2,301,591 & 384,277 & 2,301,591 & 3,104 \\
  A-Move & - & - & 22 & 22 & 22 & 22 & - & - \\
  Z-Move & 500 & 500 & 500 & 500 & - & - & - & - \\
  Merge & - & - & - & - & - & - & 15,522 & 19 \\
  Split & - & - & - & - & - & - & 0 & 0 \\ \hline
  \textbf{Proposed} & & & & & & & \\ \hline
  S & 6,156 & 6,156 & 2,156 & 2,156 & 1,156 & 1,156 & - & -\\
  Measure & 100 & 100 & 100 & 100 & 100 & 100 & 100 & - \\
  Reset & 40 & 40 & 40 & 40 & 40 & 40 & 40 & - \\
  SE & 5,151 & 5,151 & 1,151 & 1,151 & 151 & 151 & 151 & - \\
  Cult-S & 22,239 & 22,239 & 5,239 & 5,239 & 1,239 & 1,239 & 1,239 & - \\
  Cult-T & 15,149,591 & 586,077 & 1,849,591 & 351,077 & 249,591 & 251,077 & 249,591 & - \\
  Merge & - & - & - & - & - & - & 1,669 & - \\
\end{tabular}
}
\caption{Resulting Primitive operation times in based on decompositions to physical gates with speeds in Table~\ref{tab:gate-speeds} and access to resources according to Table~\ref{tab:hw-assumptions}. All times are rounded to the nearest $\mu s$. Pauli operations are considered well-defined primitives but have no time cost because they can be done virtually. Similarly, the Split primitive can always be absorbed into a preceding Merge, making its time zero in practice. For primitives depending on code distance, $d=11$ is assumed. $|T\rangle$ state cultivation includes uses the strong cultivation discard rate of $99\%$~\cite{gidney2024magicstatecultivationgrowing} for non-folded and $90\%$~\cite{acharya2022role} for folded. The Cult-S primitive is shown for movement architectures, but it is not the default behavior for the Compiler step. See Section~\ref{subsec:cult-vs-teleport} and Figure~\ref{fig:s-inject} for more details.}
\label{tab:primitive-times}
\end{table*}
\clearpage

\section{Layout Generation}
\label{app:layout-generation}
\begin{figure}
    \centering
    \includegraphics[width=\linewidth]{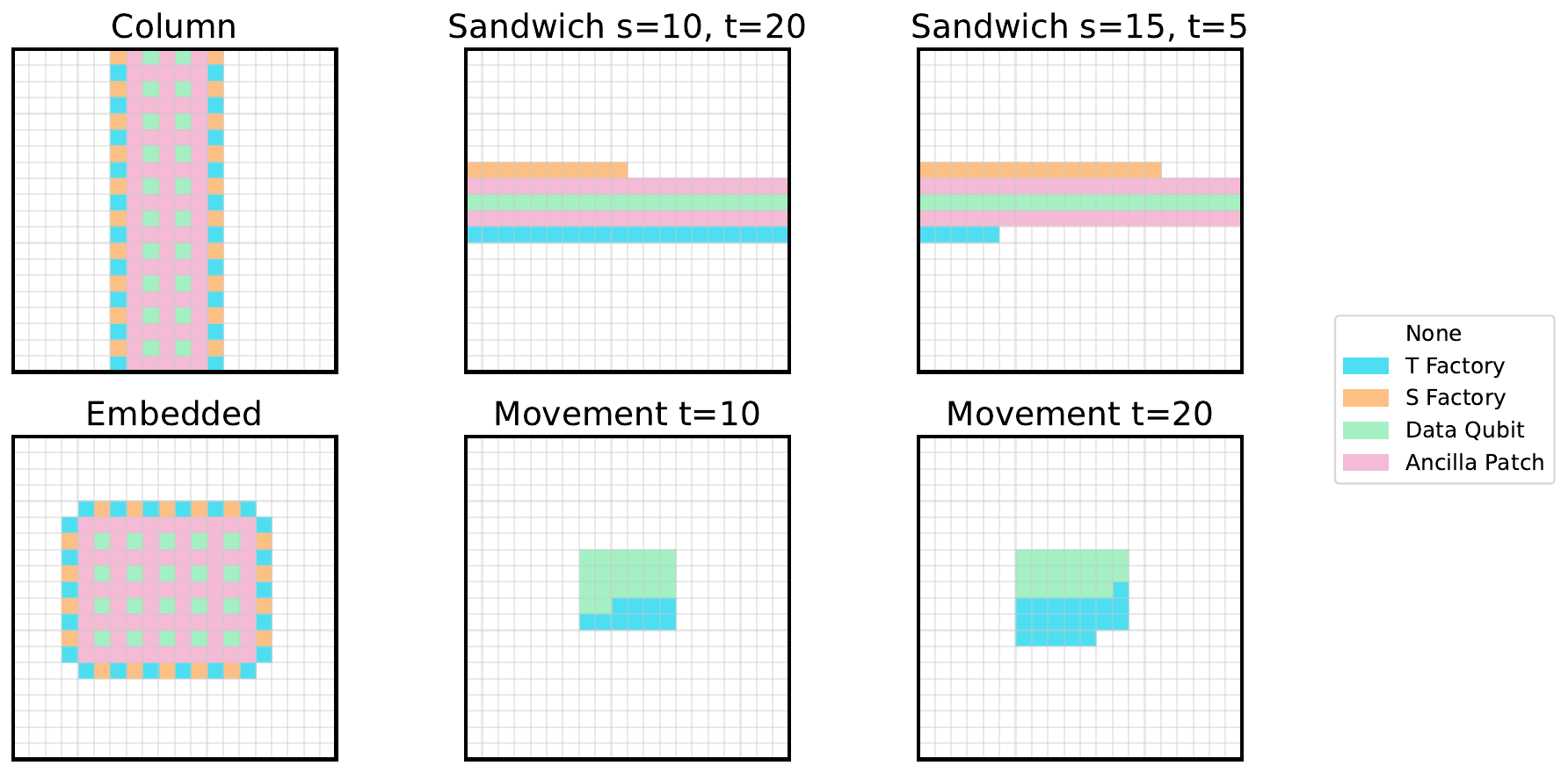}
    \caption{Example generated Layouts for both lattice surgery and movement Architectures on a circuit with $20$ data qubits. Each grid cell represents a surface code qubit patch. The Layouts in the leftmost column have a fixed number of qubits dedicated to state production, while the Layouts on the right two columns use configurable numbers of factories.}
    \label{fig:layouts}
\end{figure}
\quad The Layout is a sub-component of the larger Architecture class. Its role is to place logical qubit patches and assign those patches with individual roles. For Architectures with movement, there are only two roles qubit patches can take. A qubit patch can be either a logical qubit from the original circuit or a factory qubit added to produce $\ket{T}$ states. For Architectures using lattice surgery, two more types of qubits are necessary. These Architectures need qubit space for preparing $\ket{S}$ states, and they need ancilla patches to facilitate long range entanglement. The need for ancillas to route logical entangling operations between distant qubits is the main reason that the qubit counts for lattice surgery architectures are much higher in the main text.

For movement Architectures, our strategy is to place logical qubits densely in a square to minimize overall distance between the underlying physical qubits. Since our model treats movement between zones (Z-Move) as a fixed cost, the specific placement in the packing of logical qubits only impacts the MZO and DSM Architectures.

For lattice surgery Architectures, we consider three strategies. The ``Column'' strategy entails placing qubits from the original circuit in two columns, where each logical qubit has two factory qubits to supply resource states as close as possible. Layouts using this strategy can be procedurally generated without any further tuning of parameters. The structure facilitates parallelism, which can be useful for CNOT routing; however, the limited availability of resource states and mostly one-dimensional nature if the Layout can result in long circuit times if distant qubits need to interact. The ``Embedded'' strategy tries to be more efficient in the qubit packing portion. While interactions between data qubits are easier, the middle qubits are isolated from resource states, making it a poor choice in the environment bottlenecked by cultivation. This two-dimensional approach might be significantly improved with a dedicated mapping and routing compiler pass that could place factories in more central locations in the Layout. The last option considered is the ``Sandwich'' method of Layout generation. The key idea is to have two ancilla highways separating a line of centrally located data qubits, with each highway next to a line of factories. This method keeps open the possibility of some parallelism, while also allowing enough flexibility to include only as many factories as necessary. We find that of the three, this method tends to perform the best for circuits up to the $100$s of logical qubits and that it often approaches the cost of the other two as more factory qubits are added. For these reasons it is the default Layout in the main text.

While none of these strategies is necessarily optimal for the logical circuits considered in the main text, the ability to procedurally generate them with several degrees of control is a valuable tool for building intuition and test hypotheses about the impacts of logical qubit placement on final quantum resources. Building upon this basic framework to include more granular models of movement and more well-optimized placement strategies is left as a subject of future work.

\section{Gate Speeds}
\label{app:gate-speeds}
\quad This section provides more detail on the numbers populating Table~\ref{tab:gate-speeds}. Using available literature to derive speeds for gates at a higher layer of abstraction from the underlying physical processes can be challenging. Single and two qubit gates typically come with well-defined gate times, but other peripheral operations necessary for our model can be more ambiguous. Therefore, some simplifications must be made in order to have a well-defined model. We detail these simplifications for Reset, Measurement and Movement.
\subsection{The Reset Operation}
\quad The Reset operation is a physical operation that restores a physical qubit to the computational $\ket{0}$ (or $\ket{1}$). We associate the operation with a fidelity and an operation time. In physical systems, the Reset operation could entail one or more of the following possibilities: reloading a lost atom to the array, optically re-pumping an atom to the ground (or first excited) state, or applying a simple Pauli update on a measured physical qubit. There is evidence that doing the first can take between several hundred $\mu s$ and several thousand. Ref.~\cite{harvard-arch} does not say how long this operation takes, but were we to use the initialization time in Re.~\cite{Endres2016-ok}, we would get $0.4$ milliseconds. On the other hand, recent work~\cite{v7ny-fg31} has claimed that it is possible to load at a rate of 30000 qubits per second, which would be a loading rate of 3.3 qubits per microsecond. There is less variation in the optical pumping interpretation of resetting physical qubits. Many neutral atom architectures are focusing on non-destructive state selective readout~\cite{infq-arch}, which should reduce the burden of atom reloading. Ref.~\cite{harvard-arch} claims ``a few hundred $\mu s$'', and Ref.~\cite{lis2023midcircuitoperationsusingomgarchitecture} claims 3800 $\mu s$. In the third possibility, where a computational basis state is measured and no loss or leakage has occurred, the reset operation is the application of a Pauli operation to the physical qubit. This operation can be propagated forward so that it does not cost a physical operation, simply being absorbed into a future one. In the interest of being both reasonable and optimistic, we choose 400 $\mu s$ as the default reset time in our model. The lack of a clear standard for an important operation like Reset highlights the need to have a model that is easily configurable under different assumptions about its operation time.

\subsection{Measurement}
\quad The analysis in the main text uses a  default measurement time of 1000 $\mu s$. Ref~\cite{harvard-arch} does not appear to have a specific number for the measurement time; however, Ref.~\cite{Zhou_2025} gives a useful table that posits a $500\mu s$ measurement time and a $500\mu s$ decoding time.
We conservatively combine these two quantities in our baseline for the time taken in the measurement operation, yielding $1000\mu s$, which also agrees with the general cycle time of $1$ $ms$. We note that between measurement and reset, the effective cycle time will be higher than other works; however, ours is the direct results of attention to specific operation times that compose the cycle, rather than an assumption about the cycle itself. For the optimistic projection, we use Ref.~\cite{scott2025laser}.

\subsection{Movement}
\quad Choosing a single number for this operation is more challenging than Measurement and Reset because it is deeply tied to the physics neutral atoms. Nevertheless, in the interest of having a well-informed starting point, we chose $500\mu s$, based on Ref.~\cite{harvard-arch}'s claim of ``$2\mu  m$ over approximately $500\mu s$''. Initially, we supposed that every move took about this much time, and we continue to use this assumption for zoned architectures. Since any logical code block might be going in and out of the measurement/interaction zones, we use a flat cost to capture the average. Ref.~\cite{Zhou_2025} uses a formula based on Ref.~\cite{bluvstein2022}. While the formula is for physical movement of atoms going specific distances, we extract following formula for the time to move between sites:
\begin{equation}
    2\sqrt{\frac{12l}{5500\times10^{-6}}}\mu s
\end{equation},
where l is the number of sites from one atom to another. Our baseline $500\mu s$ corresponds with about 30 sites apart. For reference, since the surface code distance $d$ is about $2d$ qubits in length, 30 sites would allow for moving one $d=15$ surface code patch a physical distance of one logical unit in any direction. In this context $500\mu s$ seems quite fast. Like other operations, establishing a simple number for a fundamental operation in a logical processor suffers from the tension between the desire to have a complete description of the operation from physical first principles and the need to have a variable with a reasonable default value. Future improvements to the resource estimator should include more detailed models of movement as both an inter- and intra-block operation. Our current model assumes $500\mu s$ for zone movements (Z-Move) and time proportional to logical distance for alley moves (A-Move), ranging between a few $\mu s$ and $500 \mu s$.

\section{The Kanamori Hamiltonian}
\label{app:kanamori}
\quad The Kanamori model was originally formulated to describe ferromagnetism in transition metals. In the present day it is used as an impurity model to account for strong, local correlations of d-electrons within Green's function-based ab initio electronic structure theories that leverage dynamical mean-field theory (DMFT)~\cite{pashov2020questaal}. A natural use case for fault-tolerant quantum computers is to compute the interacting Green's function for local impurity models like the Kanamori model, a task whose complexity and runtime largely reduces to that of time evolution under the impurity Hamiltonian~\cite{jones2025dynamic}. As such, we perform resource estimates for a single, Trotter step, $e^{-i\hat{H}_{\text{imp}}dt}$ under the Kanamori impurity Hamiltonian

\begin{equation}\label{eq:kanamori}
\begin{aligned}
\hat{H}_{\text{imp}} = & \sum_{m, \sigma} (\epsilon_{m} - \mu) \hat{d}_{m\sigma}^\dagger \hat{d}_{m\sigma} + \hat{H}_{\text{int}} \\
& + \sum_{p, m, \sigma} \left( V_{pm} \hat{d}_{m\sigma}^\dagger \hat{c}_{p\sigma} + H.c. \right) 
+ \sum_{p, \sigma} \epsilon_{p} \hat{c}_{p\sigma}^\dagger \hat{c}_{p\sigma},
\end{aligned}
\end{equation}
where $\hat{d}_{m\sigma}^\dagger (\hat{d}_{m\sigma})$ is a fermionic creation (annihilation) operator on impurity site $m$ with spin $\sigma$. Likewise, $\hat{c}_{p\sigma}^\dagger (\hat{c}_{p\sigma})$ is a fermionic creation (annihilation) operator on bath site $p$ with spin $\sigma$. The parameters $\epsilon_m$, $\mu$, $V_{pm}$, and $\epsilon_p$ refer to the impurity energy levels, the chemical potential, the hybridization strength, and the bath energy levels of the model, respectively. Interactions occur only among the impurity sites with Hamiltonian

\begin{equation}\label{eq:kanamori_int}
\begin{split}
\hat{H}_{\text{int}} = & \ U \sum_{m} \hat{n}_{m\uparrow} \hat{n}_{m\downarrow} \\
&+ (U - 2J) \sum_{m > m', \sigma} \hat{n}_{m\sigma} \hat{n}_{m'\bar{\sigma}} \\
&+ (U - 3J) \sum_{m > m', \sigma} \hat{n}_{m\sigma} \hat{n}_{m'\sigma} \\
& - J \sum_{m \neq m'} \hat{d}_{m\uparrow}^\dagger \hat{d}_{m\downarrow} \hat{d}_{m'\downarrow}^\dagger \hat{d}_{m'\uparrow} \\
&+ J \sum_{m \neq m'} \hat{d}_{m\uparrow}^\dagger \hat{d}_{m\downarrow}^\dagger \hat{d}_{m'\downarrow} \hat{d}_{m'\uparrow},
\end{split}
\end{equation}

with Hubbard $U$ and exchange parameter $J$. The first three terms in Eq.~\ref{eq:kanamori_int} are different density-density interactions, while the fourth is the spin-flip term and the last is the pair-hopping term. The specific form we use for Eqs.~\ref{eq:kanamori}-\ref{eq:kanamori_int} is motivated by the material SrVO$_3$, whose three vanadium $t2g$ orbitals are crystal-field-split from its higher-lying $e_g$ orbitals. Because of this, we take the three $t2g$ orbitals as the three impurity sites. Meanwhile, Ref.~\cite{jamet2025anderson} found that around 27 bath sites were enough to well-discretize the continuous hybridization function of SrVO$_3$. Accounting for spin then leads to an impurity model on 60 qubits. While quantum computing the Green's function for SrVO$_3$ at 60 qubits may not confer quantum advantage, it does act as an important benchmark, both from a strongly correlated materials and problem complexity standpoint, that may herald proximate quantum utility in materials science.

\section{Details on Cultivation}
\label{app:cultivation}

\quad Cultivation is a procedure for producing magic states using error detection to increase and maintain a high fidelity state before growing the state into a larger code.
We have identified three main viable ways to perform cultivation on neutral atoms. There is the original proposal~\cite{gidney2024magicstatecultivationgrowing}, which respects nearest neighbor connectivity and comes complete with a code implementation. The main drawback is its use of a complicated grafting procedure that works with lattice surgery operations but is not compatible with transversal CNOT. Expanding on this proposal Ref.~\cite{claes2025cultivating} 
introduces a version that is close in terms of resources. It stays within the surface code and is compatible with transversal CNOT. Unfortunately, it does not have an associated code implementation to use. Most recently, folded cultivation~\cite{sahay2025foldtransversalsurfacecodecultivation} has offered an option using even fewer resources. It uses motion to stay within the surface code framework, like the previous version, but it makes an additional assumption of three-qubit gates to improve the expected number of attempts before success lowered by approximately an order of magnitude. It has some associated code, but some components are missing at the time of writing. Furthermore, access to multi-qubit gates will likely be a requirement for accessing the resource savings. While multi-qubit gates have been demonstrated for neutral atoms~\cite{PhysRevLett.123.170503}, implementing them in a scalable architecture will likely be a challenge.

Cultivation uses a sequence of checks in the form of measurements to grow a small code with a low code distance to a higher one by restarting whenever a check is failed. Since a small code can only tolerate errors of a low distance, higher fidelity $\ket{T}$ states require more growth before the program can know that higher weight errors have occurred. The key metric for the fidelity at the small distance (as opposed to the final code distance) is the fault distance. The fault distance is the number of errors from the noise model needed to cause a logical error without tripping any detectors, and it can represent the fidelity of the state in a way that is more decoupled from the code distance. Between all of these procedures there are typically two fault distances considered--three and five. Growing to higher distances might seem appealing at first, but because cultivation is a repeat-until-success procedure, the probability that no errors occur and the repetition can stop decreases exponentially. For this reason, there does not seem to be much interest in going beyond a fault distance of five. This scaling also puts a soft limit of the $\ket{T}$ state LER achievable by the procedure, which appears to be approximately $10^{-9}$ for the standard noise assumption of $p=0.001$.

While the top-level assumptions differentiate the three approaches to cultivation at a high level, the differences in implementations are important to grasp for the purpose of quantum resource estimation. These differences largely boil down to three components: which code will hold the initial state (injection), how the code will preserve and grow the fault distance (cultivation), and how to expand the small code to a larger one (escape). Ref.~\cite{gidney2024magicstatecultivationgrowing} uses the $2$d color code for injection and cultivation and a novel grafting procedure for escape. An important fact to highlight is that the final code is technically not in the rotated surface code. The difference is not very important to nearest-neighbor architectures because they do not have transversal CNOT, but it is a big deal for nonlocal ones where transversal CNOT is a key motivation. Transversal CNOT needs the data qubits in separate patches to align for the procedure to work. To address this, Ref.~\cite{Zhou_2025} applies a small merge and split to excise the grafted section at the cost of extra cycles. Ref.~\cite{claes2025cultivating} avoids needing to perform this procedure by starting in the rotated surface code, introducing a novel cultivation procedure, and relying on "standard surface code growth protocols" for escape. Finally, Ref.~\cite{sahay2025foldtransversalsurfacecodecultivation} begins in the rotated surface code, cultivates using GHZ checks (which utilize multi-qubit gates) in combination with moving between the rotated and unrotated surface code, and escapes by expanding the stabilizers to a larger patch like Ref.~\cite{claes2025cultivating}.These descriptions capture the difficulty of getting the official circuit representations of these procedures.

\end{document}